\RequirePackage{fix-cm}
\documentclass[smallextended]{svjour3}       
\smartqed  
\usepackage{graphicx}
\usepackage{amsmath}
\usepackage{amsfonts}
\usepackage{amssymb}
\usepackage{bm}
\usepackage{subfig}

\newcommand{\beq}{\begin{equation}}
\newcommand{\eeq}{\end{equation}}
\newcommand{\beqa}{\begin{eqnarray}}
\newcommand{\eeqa}{\end{eqnarray}}
\newcommand{\eq}[1]{Eq.~(\ref{#1})}
\newcommand{\nn}{\nonumber \\ }
\begin{document}

\title{The relevance of pion-exchange contributions {\it versus} contact terms in the  chiral effective field theory description of nucleon-nucleon scattering
}

\titlerunning{Relevance of pion-exchanges {\it vs.} contacts}        

\author{H. Alanazi        \and
        R. Machleidt 
}


\institute{H. Alanazi, R. Machleidt \at
              Department of Physics, University of Idaho, Moscow, ID 83844, USA\\
              \email{machleid@uidaho.edu}           
}

\date{Received: date / Accepted: date}

\maketitle

\begin{abstract}
The standard way to demonstrate the relevance of chiral symmetry for the $NN$ interaction is to consider higher partial waves of $NN$ scattering which are controlled entirely by chiral 
pion-exchanges (since contacts vanish).
However, in applications of $NN$-potentials to nuclear structure and reactions, the lower partial waves are the important ones, making the largest contributions.
Lower partial waves are sensitive to the short-range potential, and so, when the short-range contacts were to dominate over the chiral pion-contributions in lower partial waves, 
then the predictions from ``chiral potentials'' would have little to do with chiral symmetry. 
To address this issue, we investigate
 systematically the role of the (chiral) one- and two-pion exchanges, on the one hand, and the effect of the contacts, on the other hand, in the lower partial waves of 
  $NN$ scattering.  We are able to clearly identify the signature of chiral symmetry in lower partial waves.
Our study has also a pedagogical spin-off as it 
demonstrates in detail how the reproduction of the lower partial-wave phase shifts
comes about from the various ingredients of the theory.
\end{abstract}


\section{Introduction}
\label{sec_intro}

During the past three decades, it has been demonstrated that chiral effective field theory 
(chiral EFT) represents a powerful tool to deal with hadronic interactions at low energy (see Refs.~\cite{ME11,EHM09,HKK20} for recent reviews). 
By construction, chiral EFT is a model-independent theory with a firm connection to QCD
via the (broken) chiral symmetry of low-energy QCD. Moreover, the method is systematic in the sense that the various contributions to a particular dynamical process can be arranged as an expansion in terms of powers of the soft scale, $Q\sim m_\pi$, over the hard scale, 
$\Lambda_\chi \sim m_\rho$: $Q/\Lambda_\chi$ (where $m_\pi$ denotes the pion mass and $m_\rho$ the mass of the $\rho$-meson).
Within the chiral EFT approach, nucleon-nucleon ($NN$) interactions have been constructed
 up to fifth order of the chiral expansion (see Refs.~\cite{Pia15,Pia16,Car16,EMN17,RKE18,Eks18} 
for some recent examples).

These  chiral $NN$ potentials complemented by chiral three-nucleon forces 
 have been applied in
calculations of few-nucleon 
reactions~\cite{NRQ10,Viv13,Gol14,Gir19}, the structure of light- and medium-mass nuclei~\cite{Nav07a,Rot11,Rot12,Hag12a,Hag12b,BNV13,Her13,Hag14a,Bin14,HJP16,Sim16,Sim17,Mor18,Som19,Hop19,Rot19}, 
and infinite matter~\cite{HS10,Heb11,Cor13,Hag14b,Cor14,Sam15,MS16,Dri16,Heb15,DHS19,SM20,Jia20}---with, by and large, satisfactory results. These 
successes have been attributed to the chiral symmetry foundation of the 
potentials~\cite{MS20}.

In the chiral EFT procedure to nuclear forces, a clear dictinction is made between the long- and short-range parts of the $NN$ potential. While the long-range part is given by
one- and multi-pion exchanges, the short-range description
is very different.
Since the short-range nucleon structure cannot be resolved at the low energy scale characteristic for traditional nuclear physics, 
the short-range description consists simply of polynomials of
increasing degree, also known as contact terms, which contribute exclusively in the 
lower partial waves of the $NN$ interaction.

 Only the pion-contributions are ruled by chiral symmetry, while the contacts are based on just the usual non-relativistic invariances and have nothing to do with chiral symmetry.
Therefore, the standard way to demonstrate the relevance of chiral symmetry for the $NN$ interaction is to consider only higher partial waves to which contacts do not make contributions and that are, thus, controled entirely by chiral symmetry~\cite{KBW97,Ent15a,Ent15b}.

But, how about the lower partial waves?
  Here, the situation is different, because
  both contacts and pion-exchanges contribute.
   Thus, to obtain a clear picture of the role of chiral symmetry in the lower partial waves,
   the chiral pion-exchanges need to be disentagled from the contact contributions,
   which may not be easy.
   Nevertheless, this issue is of interest for the following reasons.

 Lower partial waves are more sensitive to the short-range potential. 
 Therefore, one may suspect that the contact contributions are dominant and simply override the pion-exchange contributions in lower partial waves. 
 
 In applications of $NN$-potentials to nuclear structure and reactions, the lower partial waves make the large contributions. Thus, if chiral symmetry would rule only the upper partial waves while  the lower partial wave were essentially  governed by the contacts, then the predictions from these ``chiral potentials'' for nuclear structure and reactions would have little to do with chiral symmetry. 
 Chiral potentials would not be much different from phenomenological
 potentials.
 
Motivated by the above concerns, the purpose of this paper is to systematically investigate the role of the (chiral) one- and two-pion exchange contributions, on the one hand, and the effect of the contacts, on the other hand, in the lower partial waves of chiral $NN$ potentials; to determine if chiral symmetry plays an essential role in those lower waves.

Besides this main physical motivation for this study, we like to point out that
there is  also a pedagogical spin-off to this work.
We will demonstrate in detail how the reproduction of the lower partial-wave phase shifts
comes about from the various ingredients of the theory.

In Sect.~2, will present the formalism for the chiral $NN$ potential, order by order, as it matters
to the investigation of the issue, which is conducted in Sect.~3. Sect.~4 concludes the paper.

\section{The chiral $NN$ potential}
\label{sec_pot}

\subsection{Power counting}
\label{sec_pow}

In nuclear EFT, contributions are analyzed in terms of powers of small external momenta over the large scale: $(Q/\Lambda_b)^\nu$, where $Q$ is generic for an external momentum (nucleon three-momentum or pion four-momentum) or a pion mass and $\Lambda_b \sim 0.6 $ GeV is the  breakdown scale appropriate specifically for the $NN$ problem~\cite{EKM15}. Determining the power $\nu$ has become known as power counting. Applying (naive) dimensional analysis or Weinberg counting~\cite{Wei90}, one obtains:
 for the power of a connected irreducible diagram involving $A$ nucleons~\cite{ME11},
\begin{equation}
    \nu = -2+2A-2C+2L+\sum_i {\Delta_i} ,
\end{equation}
with
\begin{equation}
    \Delta_i\equiv d_i + \frac{n_i}{2}-2 ,
\end{equation}
where $C$ denotes the number of separately connected pieces and $L$ the number of loops in the diagram; $d_i$ is the number of derivatives or pion-mass insertions and $n_i$ the number of nucleon fields (nucleon legs) involved in vertex $i$; the sum runs over all vertexes $i$ contained in the diagram under consideration. Note that $\Delta_i\ge0$ for all interactions allowed by chiral symmetry.

Since we use the heavy-baryon formalism, we encounter terms which include factors of
$Q/M_N$, where $M_N$ denotes the nucleon mass.
We count the order of such terms by the rule
$Q/M_N \sim (Q/\Lambda_\chi)^2$,
for reasons explained in Ref.~\cite{Wei90}.

In this paper, we are mainly concerned with the $NN$ system $(A=2,C=~1)$, in which case
 the power formula collapses to the very simple expression
\begin{equation}
    \nu=2L+\sum_i\Delta_i .
    \label{eq_nu}
\end{equation}

\subsection{The long-range $NN$ potential} 
\label{sec_long}

The long-range part of the $NN$ potential is built up from pion exchanges,
which are ruled by chiral symmetry.
The various pion-exchange contributions may be analyzed
according to the number of pions being exchanged between the two
nucleons:
\begin{equation}
V_{\pi}= V_{1\pi}+V_{2\pi}+V_{3\pi}+\cdots,
\end{equation}
 where the meaning of the subscripts is obvious and the ellipsis represents $4\pi$ and higher pion exchanges. For each of the above terms, we have a low-momentum expansion: 
 \begin{align}
  V_{1\pi} &=V_{1\pi}^{(0)}+V_{1\pi}^{(2)}+V_{1\pi}^{(3)}+V_{1\pi}^{(4)}+V_{1\pi}^{(5)}+\cdots , \\
  V_{2\pi} &=V_{2\pi}^{(2)}+V_{2\pi}^{(3)}+V_{2\pi}^{(4)}+V_{2\pi}^{(5)}+\cdots , \\
  V_{3\pi} &=V_{3\pi}^{(4)}+V_{3\pi}^{(5)}+\cdots,
 \end{align}
where the superscript denotes the order $\nu$ of the expansion.

 Order by order, the long-range $NN$ potential builds up as follows:
 \begin{align}
     V_{L\Omega} &\equiv  V^{(0)}=V_{1\pi}^{(0)} ,\\
     V_{NL\Omega} &\equiv V^{(2)}=V_{L\Omega}+V_{1\pi}^{(2)}+V_{2\pi}^{(2)}, \\
     V_{NNL\Omega} &\equiv  V^{(3)}=V_{NL\Omega}+V_{1\pi}^{(3)}+V_{2\pi}^{(3)} ,\\
      V_{N^3L\Omega} &\equiv  V^{(4)}=V_{NNL\Omega}+V_{1\pi}^{(4)}+V_{2\pi}^{(4)}+V_{3\pi}^{(4)} ,\\
       V_{N^4L\Omega} &\equiv  V^{(5)}=V_{N^3L\Omega}+V_{1\pi}^{(5)}+V_{2\pi}^{(5)}+V_{3\pi}^{(5)},
 \end{align}
where L$\Omega$ stands for leading order, NL$\Omega$ for next-to-leading order, etc..

\subsubsection{Leading order}
\label{sec_lo}

\begin{table}[t]
\caption{Basic constants used throughout this work~\cite{PDG}.}
\label{tab_basic}
\smallskip
\begin{tabular*}{\textwidth}{@{\extracolsep{\fill}}lcl}
\hline 
\hline 
\noalign{\smallskip}
  Quantity            
 &  \hspace{2cm} 
 & Value \\
\hline
\noalign{\smallskip}
Axial-vector coupling constant $g_A$ && 1.29 \\
Pion-decay constant $f_\pi$ && 92.4 MeV \\
Charged-pion mass $m_{\pi^\pm}$ && 139.5702 MeV \\
Neutral-pion mass $m_{\pi^0}$ && 134.9766 MeV \\
Average pion-mass $\bar{m}_\pi$ && 138.0390 MeV \\
Proton mass $M_p$ && 938.2720 MeV \\
Neutron mass $M_n$ && 939.5654 MeV \\
Average nucleon-mass $\bar{M}_N$ && 938.9183 MeV \\
\hline
\hline
\noalign{\smallskip}
\end{tabular*}
\end{table}

At leading order (L$\Omega$, $\nu=0$), only one-pion exchange (1PE) contributes to the long range. The charge-independent 1PE is given by 
\begin{equation}
    V_{1\pi}^{(CI)}(\Vec{p'},\Vec{p})= -\:\frac{g_{A}^2}{4f_{\pi}^2}\: \bm{\tau}_1 \cdot \bm{\tau}_2\:   
    \frac{{\Vec\sigma}_1 \cdot \Vec{q} \:\:  \Vec\sigma_2 \cdot \Vec{q}}{q^2+m_{\pi}^2}\: ,
    \label{eq_1peci}
\end{equation}
where $\Vec{p'}$ and $\Vec{p}$ denote the final and initial nucleon momenta in the center-of-mass system (CMS), respectively. Moreover, $\vec q = {\vec p}\,' - \vec p$ is the momentum transfer and $\Vec\sigma_{1,2}$ and $\bm{\tau}_{1,2}$ are the spin and isospin operators of nucleons 1 and 2. Parameters $g_A, f_\pi$ and $m_\pi$ denote the axial-vector coupling constant, pion-decay constant, and the pion mass, respectively.
See Table~\ref{tab_basic} for their values.
Higher order corrections to the 1PE  are taken care of by  mass
and coupling constant renormalizations. Note also that, on 
shell, there are no relativistic corrections. Thus, we apply  1PE in the form
\eq{eq_1peci} through all orders.

The 1PE potential, Eq.~(\ref{eq_1peci}), can be re-written as follows:
\begin{equation}
    V_{1\pi}^{(CI)}(\Vec{p'},\Vec{p})=
     -\:\frac{g_{A}^2}{12f_{\pi}^2}\: \bm{\tau}_1 \cdot \bm{\tau}_2
    \left( 
    \Vec\sigma_1 \cdot \Vec\sigma_2 \,
    -\, \Vec\sigma_1 \cdot \Vec\sigma_2 \,
      \frac{m_{\pi}^2}{q^2+m_{\pi}^2}
    +\,  \frac{S_{12}(\Vec{q})}{q^2+m_{\pi}^2}\:
      \right),
    \label{eq_1pe_alt}
\end{equation}
with tensor operator
\begin{equation}
S_{12}(\Vec{q}) = 3 \: \Vec\sigma_1 \cdot \Vec{q} \:\:  \Vec\sigma_2 \cdot \Vec{q} \,
  - \, \Vec\sigma_1 \cdot \Vec\sigma_2 \, q^2 \,,
\end{equation}
where the 1PE has been broken up into a zeroth order spin-spin contact term (``$\delta$-function term''), a spin-spin Yukawa central force,
and a tensor piece. The 1PE tensor force is known to be strong, while the spin-spin central force
is weak.

If one takes the charge-dependence of the 1PE into account, then, in proton-proton ($pp$) 
and neutron-neutron ($nn$) scattering one has
\begin{equation}
  V_{1\pi}^{(pp)}(\Vec{p'},\Vec{p})=  V_{1\pi}^{(nn)}(\Vec{p'},\Vec{p})= V_{1\pi}(m_{\pi^0})
\end{equation}
and in $np$ scattering
\begin{equation}
   V_{1\pi}^{(np)}(\Vec{p'},\Vec{p})= -  V_{1\pi}(m_{\pi^0}) + (-1)^{I+1}\:2 \:V_{1 \pi}(m_{\pi^\pm})\:, 
   \label{eq_1pe_np}
\end{equation}
where $I=0,1$ denotes the total isospin of the two-nucleon system and 
\begin{equation}
    V_{1\pi}(m_\pi) = -\:\frac{g_{A}^2}{4f_{\pi}^2}\:    \frac{\Vec\sigma_1 \cdot \Vec{q}\:  \Vec\sigma_2 \cdot  \Vec{q}}{q^2+m_{\pi}^2}\:.  
\end{equation}
The charge-dependence of the 1PE is of order NL$\Omega$~\cite{ME11}, 
but we include it already at order L$\Omega$ to make the comparison with the $np$ phase-shift analyses meaningful.

\subsubsection{Next-to-leading order}
\label{sec_nlo}

At next-to-leading order (NL$\Omega$, $\nu=2$), two-pion exchange (2PE) starts  and continues through all higher orders.
The 2PE potential expressions will be stated in terms of contributions to the 
momentum-space $NN$ amplitudes in the CMS, 
which takes the following general form:
\begin{eqnarray} 
V({\vec p}~', \vec p) &  = &
 \:\, V_C \:\, + \bm{\tau}_1 \cdot \bm{\tau}_2 \, W_C 
\nonumber \\ & +&  
\left[ \, V_S \:\, + \bm{\tau}_1 \cdot \bm{\tau}_2 \, W_S 
\,\:\, \right] \,
\vec\sigma_1 \cdot \vec \sigma_2
\nonumber \\ &+& 
\left[ \, V_{LS} + \bm{\tau}_1 \cdot \bm{\tau}_2 \, W_{LS}    
\right] \,
\left(-i \vec S \cdot (\vec q \times \vec k) \,\right)
\nonumber \\ &+& 
\left[ \, V_T \:\,     + \bm{\tau}_1 \cdot \bm{\tau}_2 \, W_T 
\,\:\, \right] \,
\vec \sigma_1 \cdot \vec q \,\, \vec \sigma_2 \cdot \vec q  
\nonumber \\ &+& 
\left[ \, V_{\sigma L} + \bm{\tau}_1 \cdot \bm{\tau}_2 \, 
      W_{\sigma L} \, \right] \,
\vec\sigma_1\cdot(\vec q\times \vec k\,) \,\,
\vec \sigma_2 \cdot(\vec q\times \vec k\,)
\, ,
\label{eq_nnamp}
\end{eqnarray}
where
$\vec k =({\vec p}\,' + \vec p)/2$ is the average momentum and $\vec S =(\vec\sigma_1+
\vec\sigma_2)/2 $ the total spin.
For on-shell scattering, $V_\alpha$ and $W_\alpha$ ($\alpha=C,S,LS,T,\sigma L$) can be 
expressed as functions of $q= |\vec q\,|$ and $p=|{\vec p}\,'| = |\vec p\,|$, only.

The  NL$\Omega$ 2PE is given by~\cite{KBW97,Ent15a}:
\begin{eqnarray} 
W_C &=&{L(\tilde{\Lambda};q)\over384\pi^2 f_\pi^4} \left[4m_\pi^2(1+4g_A^2-5g_A^4)
+q^2(1+10g_A^2-23g_A^4) - {48g_A^4 m_\pi^4 \over w^2} \right] 
\nonumber \\
 &&  + \,  \mbox{\rm polynomial terms of order two} 
\,,  
\label{eq_2C}
\\   
V_T &=& -{1\over q^2} V_{S} \; = \; -{3g_A^4 \over 64\pi^2 f_\pi^4} L(\tilde{\Lambda};q)
 \, + \, \mbox{\rm polynomial terms of order zero} 
\,,
\label{eq_2T}
\end{eqnarray}  
with $ w = \sqrt{4m_\pi^2+q^2}$ and
where the (regularized) logarithmic loop function is given by
\begin{equation} 
L(\tilde{\Lambda};q) = {w\over 2q} 
\ln {\frac{\tilde{\Lambda}^2(2m_\pi^2+q^2)-2m_\pi^2 q^2+\tilde{\Lambda}\sqrt{
\tilde{\Lambda}^2-4m_\pi^2}\, q\,w}{2m_\pi^2(\tilde{\Lambda}^2+q^2)}} \,.
\label{eq_L}
\end{equation}
Note that
\begin{equation}
\lim_{\tilde{\Lambda} \rightarrow \infty} L(\tilde{\Lambda};q) =  {w\over q} 
\ln {\frac{w+q}{2m_\pi}} 
\end {equation}
results in the logarithmic loop function of dimensional regularization. 

For the explicit expressions of the polynomial terms that contribute to
Eqs.~(\ref{eq_2C}) and (\ref{eq_2T}), see Ref.~\cite{KBW97}.

\subsubsection{Next-to-next-to-leading order}
\label{sec_nnlo}

 \begin{table}\centering
 \caption{Values of the $\pi N$ LECs as determined in Ref.~\cite{Hof15}. The $c_i$ and $\Bar{d_i}$ are the LECs of the second and third order $\pi N$ Lagrangians and are in units of GeV$^{-1}$ and GeV$^{-2}$, respectively.
 The uncertainties in the last digit are given in parentheses after the value.}
    \smallskip
    \begin{tabular*}{\textwidth}{@{\extracolsep{\fill}}crr}
    \hline \hline
      \noalign{\smallskip}
         &  NNL$\Omega$ & N$^3$L$\Omega$  \\
      \hline 
        \noalign{\smallskip}
        $c_1$&-0.74(2)&-1.07(2) \\
        $c_2$&&3.20(3) \\
        $c_3$&-3.61(5)&-5.32(5)\\
        $c_4$& 2.44(3)&3.56(3)\\
        $\Bar{d_1}+\Bar{d_2}$&&1.04(6)\\
        $\Bar{d_3}$&&-0.48(2)\\
        $\Bar{d_5}$&&0.14(5)\\
        $\Bar{d}_{14}-\Bar{d}_{15}$&&-1.90(6)\\
         \noalign{\smallskip}
     \hline \hline
\end{tabular*}
   \label{tab_LEC_piN}
\end{table}

At next-to-next-to-leading order (NNL$\Omega$, $\nu=3$), we have the following
 2PE  contributions~\cite{KBW97,Ent15a}:
\begin{eqnarray} 
V_C &=&  {3g_A^2 \over 16\pi f_\pi^4} \left[2m_\pi^2(c_3- 2c_1)+c_3 q^2 \right](2m_\pi^2+q^2) 
A(\tilde{\Lambda};q) 
\nonumber \\
 &&  + \,  \mbox{\rm polynomial terms of order three} 
\,, 
\label{eq_3C}
\\
W_T &=&-{1\over q^2}W_{S} =-{g_A^2 \over 32\pi f_\pi^4} c_4 w^2  A(\tilde{\Lambda};q)
\nonumber \\ &&  
 + \, \mbox{\rm polynomial terms of order one} 
\,.
\label{eq_3T}
\end{eqnarray}   
The loop function that appears in the above expressions,
regularized by spectral-function cut-off $\tilde{\Lambda}$, is
\begin{equation} 
A(\tilde{\Lambda};q) = {1\over 2q} \arctan{q (\tilde{\Lambda}-2m_\pi) \over q^2
+2\tilde{\Lambda} m_\pi} \,,
\label{eq_A}
\end{equation}
with
\begin{equation}
\lim_{\tilde{\Lambda} \rightarrow \infty} A(\tilde{\Lambda};q) =  
{1\over 2q} \arctan{q \over 2m_\pi} 
\end {equation}
yielding the corresponding loop function of dimensional regularization.

The polynomial terms that occur in
Eqs.~(\ref{eq_3C}) and (\ref{eq_3T}) are given in Ref.~\cite{KBW97}.

In the above expressions, the $\pi N$
low-energy constants (LECs), $c_i$, from the second order $\pi N$ Lagrangian
appear for the first time. We use the very precise values as determined in the Roy-Steiner analysis 
by Ref.~\cite{Hof15} (Table~\ref{tab_LEC_piN}).

\subsubsection{Next-to-next-to-next-to-leading order}
\label{sec_n3lo}

At next-to-next-to-next-to-leading order (N$^3$L$\Omega$, $\nu=4$),
we have many contributions which we will subdivide into separate groups.

\paragraph{Football diagram at N$^3$L$\Omega$.}
The 2PE football diagram at N$^3$L$\Omega$ generates~\cite{Kai01}:
\begin{eqnarray}  
V_C & = & {3\over 16 \pi^2 f_\pi^4 } 
\left[\left( {c_2 \over 6} w^2 +c_3(2m_\pi^2+q^2) -4c_1 m_\pi^2 \right)^2 
+{c_2^2 \over 45 } w^4 \right]  L(\tilde{\Lambda};q) \,, 
\label{eq_4c2C}
\\
W_T  &=&  -{1\over q^2} W_S 
     = {c_4^2 \over 96 \pi^2 f_\pi^4 }  w^2 L(\tilde{\Lambda};q)
\,.
\label{eq_4c2T}
\end{eqnarray}
In addition to the non-polynomial contributions shown in the above equations,
there are polynomial terms of order four in the central potential and
polynomial terms of order two in the tensor (and spin-orbit) potentials,
which we do not show explicitly. This note applies to all N$^3$L$\Omega$
expressions.

\paragraph{Leading two-loop contributions.}

We state the 2PE two-loop contributions in terms of their
spectral functions~\cite{Kai01}:
\begin{eqnarray} 
{\rm Im}\, V_C &=& 
{3g_A^4 (2m_\pi^2-\mu^2) \over \pi \mu (4f_\pi)^6} 
\Bigg[ 
(m_\pi^2-2\mu^2) 
\left( 
2m_\pi +{2m_\pi^2 -\mu^2 \over2\mu} \ln{\mu+2m_\pi \over \mu-2m_\pi} 
\right) 
\nonumber \\ && 
+ \, 4g_A^2 m_\pi(2m_\pi^2-\mu^2) 
\Big] 
\,,
\\
{\rm Im}\, W_C &=& 
{2\kappa \over 3\mu (8\pi f_\pi^2)^3} 
\int_0^1 dx\, 
\Big[ g_A^2(\mu^2-2m_\pi^2) +2(1-g_A^2)\kappa^2x^2 \Big]
\nonumber \\ && \times 
\Bigg\{
\,96 \pi^2 f_\pi^2 
\left[ 
(2m_\pi^2-\mu^2)(\bar{d}_1 +\bar{d}_2) -2\kappa^2x^2 \bar{d}_3+4m_\pi^2 \bar{d}_5 
\right] 
\nonumber \\ && +
\left[ 
4m_\pi^2 (1+2g_A^2) -\mu^2(1+5g_A^2)
\right] 
{\kappa\over \mu} \ln {\mu +2\kappa\over 2m_\pi} \,
  \nonumber \\ && 
+\,{\mu^2 \over 12} (5+13g_A^2) -2m_\pi^2 (1+2g_A^2) 
\nonumber \\ && 
-\,3\kappa^2x^2 +6 \kappa x \sqrt{m_\pi^2 +\kappa^2 x^2} \ln{ \kappa x +\sqrt{m_\pi^2 
+\kappa^2 x^2}\over  m_\pi}
\nonumber \\ && 
+ \, g_A^4\left(\mu^2 -2\kappa^2 x^2 -2m_\pi^2\right) 
\nonumber \\ && \times
\left[ 
{5\over 6} +{m_\pi^2\over \kappa^2 x^2} 
-\left( 1 +{m_\pi^2\over \kappa^2 x^2} \right)^{3/2} 
\ln{ \kappa x +\sqrt{m_\pi^2 +\kappa^2 x^2}\over  m_\pi} 
\right] 
\Bigg\} 
\,,   
\\
{\rm Im}\, V_S &=&
\mu^2\,{\rm Im}\, V_T = 
{g_A^2\mu \kappa^3 \over 8\pi f_\pi^4} \left(\bar{d}_{15}-\bar{d}_{14}\right) 
\, + \, {2g_A^6\mu \kappa^3 \over (8\pi f_\pi^2)^3}
 \int_0^1 dx(1-x^2)
 \nonumber \\ && \times
 \left[ 
 {1\over 6}-{m_\pi^2 \over \kappa^2x^2} 
+\left( 
1+{m_\pi^2 \over \kappa^2x^2} 
\right)^{3/2} 
\ln{ \kappa x +\sqrt{m_\pi^2 +\kappa^2 x^2}\over  m_\pi}
\right] 
\,,
\\
{\rm Im}\, W_S &=& 
\mu^2 \,{\rm Im}\, W_T(i\mu) = 
{g_A^4(4m_\pi^2-\mu^2) \over \pi (4f_\pi)^6} 
 \nonumber \\ && \times
\left[ 
\left( 
m_\pi^2 -{\mu^2 \over 4} 
\right)
\ln{\mu+2m_\pi \over \mu-2m_\pi} 
+(1+2g_A^2)\mu  m_\pi
\right]
\,, 
\end{eqnarray}
where $\kappa = \sqrt{\mu^2/4-m_\pi^2}$.
\normalsize
The above expressions involve the $\pi N$ LECs, $\bar{d}_i$, from the third order $\pi N$ Lagrangian.
The values we apply for these LECs are shown in Table~\ref{tab_LEC_piN}.

The momentum space amplitudes $V_\alpha(q)$ and $W_\alpha(q)$
are obtained from the above spectral functions
via the subtracted dispersion integrals,
\begin{eqnarray} 
V_{C,S}(q) &=& 
-{2 q^6 \over \pi} \int_{2m_\pi}^{\tilde{\Lambda}} d\mu \,
{{\rm Im\,}V_{C,S}(i \mu) \over \mu^5 (\mu^2+q^2) }\,, 
\nn
V_T(q) &=& 
{2 q^4 \over \pi} \int_{2m_\pi}^{\tilde{\Lambda}} d\mu \,
{{\rm Im\,}V_T(i \mu) \over \mu^3 (\mu^2+q^2) }\,, 
\label{eq_disp}
\end{eqnarray}
and similarly for $W_{C,S,T}$.
For $\tilde{\Lambda} \rightarrow \infty$ the above dispersion integrals yield the
results of dimensional regularization, while for finite $\tilde{\Lambda} \geq 2m_\pi$
we have what has become known  as spectral-function regularization (SFR) \cite{EGM04}. The 
purpose of the finite scale $\tilde{\Lambda}$ is to constrain the imaginary parts to the  
low-momentum region where chiral effective field theory is applicable.  
Thus, a reasonable choice for $\tilde{\Lambda}$ is to keep it below the masses of the vector mesons
$\rho(770)$ and $\omega(782)$, but above the $f_0(500)$ [also know as $\sigma(500)$]~\cite{PDG}.
This suggests that the region 600-700 MeV is appropriate for $\tilde{\Lambda}$.
Consequently, we use $\tilde{\Lambda} =650$ MeV.

The subtracted dispersion integrals generate (besides the non-polynomial contributions)
polynomial terms of order four for the central potentials and 
polynomial tertms of order two for the tensor potentials.

\paragraph{Leading relativistic corrections.}

The relativistic corrections to the 2PE NL$\Omega$ diagrams
count as N$^3$L$\Omega$ ($\nu=4$) and are given by~\cite{ME11}:
\begin{eqnarray}
V_C &=& \frac{3 g_A^4}{128 \pi f_\pi^4 M_N} 
\bigg[\frac{m_\pi^5}{2w^2}+(2m_\pi^2+q^2)(q^2-m_\pi^2) A(\tilde{\Lambda};q) \bigg]
\,,
\label{eq_3EM1}
\\
W_C &=& \frac{g_A^2}{64 \pi f_\pi^4 M_N} 
\Bigg\{ 
\frac{3g_A^2m_\pi^5}{2 w^2} 
+ \big[g_A^2 (3m_\pi^2+2q^2) 
- 2m_\pi^2-q^2\big] 
\nonumber \\ && \times
(2m_\pi^2+q^2) A(\tilde{\Lambda};q) 
\Bigg\}
\,,
\\
V_T &=& -\frac{1}{q^2} V_S = \frac{3 g_A^4}{256 \pi f_\pi^4 M_N} 
(5m_\pi^2+2q^2) A(\tilde{\Lambda};q)\,,
\\
W_T &=& -\frac{1}{q^2} W_S = \frac{g_A^2}{128 \pi f_\pi^4 M_N} 
\big[g_A^2 (3m_\pi^2+q^2)-w^2 \big] A(\tilde{\Lambda};q) \,,
\label{eq_3EM4}
\\
V_{LS} &=&  {3g_A^4  \over 32\pi f_\pi^4 M_N} \, (2m_\pi^2+q^2) A(\tilde{\Lambda};q)
 \,,\\  
W_{LS} &=& {g_A^2(1-g_A^2)\over 32\pi f_\pi^4 M_N} \, w^2 A(\tilde{\Lambda};q) \,.
\end{eqnarray}

\paragraph{Leading three-pion exchange}
The leading $3\pi$-exchange contributions that occur at N$^3$L$\Omega$
have been calculated in Refs.~\cite{Kai00a,Kai00b} and are found to be negligible. We, 
therefore, omit them.

\subsubsection{Subleading relativistic corrections}
\label{sec_rel}

We also include
the $1/M_N$ corrections of the 2PE NNL$\Omega$ diagrams.
This contribution is repulsive and proportional to $c_i/M_N$.
Since we count $Q/M_N \sim (Q/\Lambda_b)^2$, these relativistic 
corrections  are formally of order N$^4$L$\Omega$ ($\nu=5$), but we add them
to our N$^3$L$\Omega$ potential
to compensate the excessive attraction
generated by the football diagram at N$^3$L$\Omega$~\cite{EMN17}.

The result for this group of diagrams reads \cite{Kai01,EMN17}:
\begin{eqnarray} 
V_C & = & {g_A^2\, L(\tilde{\Lambda};q) \over 32 \pi^2 M_N f_\pi^4 } 
\Big[ 
(6c_3-c_2) q^4 +4(3c_3-c_2-6c_1)q^2 m_\pi^2
\nonumber \\ &&
+ \, 6(2c_3-c_2)m_\pi^4- 24(2c_1+c_3)m_\pi^6 w^{-2} 
\Big] 
\,,
\label{eq_4cMC}
\\
W_C &=& -{c_4 \over 192 \pi^2 M_N f_\pi^4 } 
\left[ g_A^2 (8m_\pi^2+5q^2) + w^2 \right] q^2 \,  L(\tilde{\Lambda};q)
\,, \\
W_T  &=&  -{1\over q^2} W_S = {c_4 \over 192 \pi^2 M_N f_\pi^4 } 
\left[ w^2-g_A^2 (16m_\pi^2+7q^2) \right]  L(\tilde{\Lambda};q)
\label{eq_4cMS}
\,,  \\
V_{LS}& = & {c_2 \, g_A^2 \over 8 \pi^2 M_N f_\pi^4 } 
\, w^2 L(\tilde{\Lambda};q) 
\,, \\
W_{LS}  &=& 
-{c_4  \over 48 \pi^2 M_N f_\pi^4 } 
\left[ g_A^2 (8m_\pi^2+5q^2) + w^2 \right]  L(\tilde{\Lambda};q)
\,.
\label{eq_4cMLS}
\end{eqnarray}

\subsection{The short-range $NN$ potential ($NN$ contact terms)}
\label{sec_short}

In the EFT approach, the short range interaction is described by contributions of the contact type, which are constrained by parity, time-reversal, and the usual invariances, but not by chiral symmetry. Only even powers of momentum are allowed because of parity and time-reversal. Thus, the expansion of the contact potential is formally given by 
\begin{equation}
    V_{ct}= V_{ct}^{(0)}+V_{ct}^{(2)}+V_{ct}^{(4)}+V_{ct}^{(6)}+\cdots\:,
\end{equation}
where the superscript denotes the power or order.

In operator form, the contact potentials are given by:\\
Zeroth-order (leading order, L$\Omega$),
\begin{equation}
    V_{ct}^{(0)}(\Vec{p'},\Vec{p})= C_S+C_T\: \Vec\sigma_1\cdot\Vec\sigma_2 \,.
       \label{eq_ct0}
\end{equation}
Second order (next-to-leading order, NL$\Omega$),
\begin{align}
    V_{ct}^{(2)}(\Vec{p'},\Vec{p})&=C_1 q^2+C_2 k^2+(C_3 q^2+C_4 k^2)\Vec{\sigma_1}\cdot\Vec{\sigma_2}\nonumber\\
    &+C_5[-i\Vec{S}\cdot(\Vec{q}\times\Vec{k})]+C_6 (\Vec{\sigma_1}\cdot\Vec{q})(\Vec{\sigma_2}\cdot\Vec{q})\nonumber\\
    &+C_7(\Vec{\sigma_1}\cdot\Vec{k})(\Vec{\sigma_2}\cdot\Vec{k}) \,.
      \label{eq_ct2}
\end{align}
Fourth order (next-to-next-to-next-to-leading order, N$^3$L$\Omega$):
\begin{align}
  V_{ct}^{(4)}(\Vec{p'},\Vec{p})&= D_1q^4+D_2k^4+D_3q^2k^2+D_4(\Vec{q}\times\Vec{k} )^2\nonumber\\
  &+[D_5q^4+D_6k^4+D_7q^2k^2+D_8(\Vec{q}\times\Vec{k})^2]\Vec{\sigma_1}\cdot\Vec{\sigma_2}\nonumber\\
  &+(D_9q^2+D_{10}k^2)[-i\Vec{S}\cdot(\Vec{q}\times\Vec{k}]\nonumber\\
  &+(D_{11}q^2+D_{12}k^2)(\Vec{\sigma_1}\cdot\Vec{q})(\Vec{\sigma_2}\cdot\Vec{q})\nonumber\\
   &+(D_{13}q^2+D_{14}k^2)(\Vec{\sigma_1}\cdot\Vec{k})(\Vec{\sigma_2}\cdot\Vec{k})\nonumber\\
   &+D_{15}[\Vec{\sigma_1}\cdot(\Vec{q}\times\Vec{k})\Vec{\sigma_2}\cdot(\Vec{q}\times\Vec{k})] \,.
     \label{eq_ct4}
 \end{align}

In terms of a partial-wave decomposition, we have up to fourth order:
 \begin{align}
\langle {}^{1}S_0, p' |  V_{ct} | {}^{1}S_0, p \rangle &=\widetilde{C}_{{}^{1}S_0}+C_{{}^{1}S_0}(p^2+p'^2)+\widehat{D}_{{}^{1}S_0}(p'^4+p^4) + {D}_{{}^{1}S_0}p'^2p^2 ,\nonumber\\
  \langle {}^{3}S_1, p' |  V_{ct} | {}^{3}S_1, p \rangle &=\widetilde{C}_{{}^{3}S_1}+C_{{}^{3}S_1}(p^2+p'^2)+\widehat{D}_{{}^{3}S_1}(p'^4+p^4)+{D}_{{}^{3}S_1} p'^2 p^2,\nonumber\\
    \langle {}^{3}S_1, p' |  V_{ct} | {}^{3}D_1, p \rangle    &=C_{{}^{3}S_1-{}^{3}D_1}p^2+\widehat{D}_{{}^{3}S_1-{}^{3}D_1}p^4+ {D}_{{}^{3}S_1-{}^{3}D_1}p'^2 p^2,\nonumber\\
 \langle {}^{1}P_1, p' |  V_{ct} | {}^{1}P_1, p \rangle &=C_{{}^{1}P_1}\:pp'+D_{{}^{1}P_1}(p'^3p+p'p^3),\nonumber\\
  \langle {}^{3}P_0, p' |  V_{ct} | {}^{3}P_0, p \rangle  &=C_{{}^{3}P_0}\:pp'+D_{{}^{3}P_0}(p'^3p+p'p^3),\nonumber\\
   \langle {}^{3}P_1, p' |  V_{ct} | {}^{3}P_1, p \rangle &=C_{{}^{3}P_1}\:pp'+D_{{}^{3}P_1}(p'^3p+p'p^3),\nonumber\\
  \langle {}^{3}P_2, p' |  V_{ct} | {}^{3}P_2, p \rangle  &=C_{{}^{3}P_2}\:pp'+D_{{}^{3}P_2}(p'^3p+p'p^3),\nonumber\\
     \langle {}^{3}P_2, p' |  V_{ct} | {}^{3}F_2, p \rangle     &= {D}_{{}^{3}P_2-{}^{3}F_2}p'p^3, \nonumber\\
    \langle {}^{1}D_2, p' |  V_{ct} | {}^{1}D_2, p \rangle  &={D}_{{}^{1}D_2} p'^2 p^2,\nonumber\\
     \langle {}^{3}D_1, p' |  V_{ct} | {}^{3}D_1, p \rangle  &={D}_{{}^{3}D_1} p'^2 p^2,\nonumber\\
       \langle {}^{3}D_2, p' |  V_{ct} | {}^{3}D_2, p \rangle  &={D}_{{}^{3}D_2} p'^2 p^2,\nonumber\\
       \langle {}^{3}D_3, p' |  V_{ct} | {}^{3}D_3, p \rangle  &={D}_{{}^{3}D_3} p'^2 p^2.
            \label{eq_ct}
\end{align}

Notice that, in our notation, partial-wave contact LECs
\begin{itemize}
\item
$\widetilde{C}_\alpha$ are of zeroth order (there are two),
\item
${C}_\alpha$ are of  second order (there are seven), and
\item
$\widehat{D}_\alpha$ and ${D}_\alpha$ are of fourth order (there are 15),
\end{itemize}
where $\alpha$ stands for a partial wave or a combination thereof.
There exist linear one-to-one relations between the two $\widetilde{C}_\alpha$ and $C_S$ and $C_T$ of Eq.~(\ref{eq_ct0}), the seven $C_\alpha$ and the seven $C_i$ of Eq.~(\ref{eq_ct2}), and
the 15 $\widehat{D}_\alpha$ and  $D_\alpha$ and the 15 $D_i$ of Eq.~(\ref{eq_ct4}). 
The relations can be found in Ref.~\cite{ME11}.

Note that the partial-wave decomposition of
$Q^\nu$ 
(where $Q$ is either
the momentum transfer $q$ or the average momentum $k$)
has an interesting property.
For even $\nu$,
\begin{equation}
Q^\nu = 
f_{\frac{\nu}{2}}(\cos \theta) 
\, ,
\end{equation}
where $f_m$ stands for a polynomial of degree $m$
and $\theta$ is the CMS scattering angle.
The partial-wave decomposition of $Q^\nu$ for a state
of orbital-angular momentum $L$
involves the integral
\begin{equation}
I^{(\nu)}_L  
=\int_{-1}^{+1} Q^\nu P_L(\cos \theta) d\cos \theta 
=\int_{-1}^{+1}
f_{\frac{\nu}{2}}(\cos \theta) 
 P_L(\cos \theta) d\cos \theta 
\,,
\end{equation}
where $P_L$ is a Legendre polynomial.
Due to the orthogonality of the $P_L$, 
\begin{equation}
I^{(\nu)}_L = 0  
\hspace*{.5cm}
\mbox{for}
\hspace*{.5cm}
L >\frac{\nu}{2} \, .
\label{eq_ct00}
\end{equation}
Consequently, contact terms of order zero contribute only
in $S$-waves, while second order terms can contribute up to 
$P$-waves, fourth order terms up to $D$-waves,
etc..

\subsection{Scattering equation and regularization}
\label{sec_reg}

The potential $V$ is, in principal, an invariant amplitude (with relativity taken into account perturbatively) and, thus, satisfies a relativistic scattering equation, like, e.~g., the
Blankenbeclar-Sugar (BbS) equation~\cite{BS66},
which reads explicitly,
\begin{equation}
{T}({\vec p}~',{\vec p})= {V}({\vec p}~',{\vec p})+
\int \frac{d^3p''}{(2\pi)^3} \:
{V}({\vec p}~',{\vec p}~'') \:
\frac{M_N^2}{E_{p''}} \:  
\frac{1}
{{ p}^{2}-{p''}^{2}+i\epsilon} \:
{T}({\vec p}~'',{\vec p}) \,,
\label{eq_bbs2}
\end{equation}
with $E_{p''}\equiv \sqrt{M_N^2 + {p''}^2}$.
The advantage of using a relativistic scattering equation is that it automatically
includes relativistic kinematical corrections to all orders. Thus, in the scattering equation,
no propagator modifications are necessary when moving up in the orders.

Defining
\begin{align}
    \widehat{V}(\Vec{p'},\Vec{p})\equiv\frac{1}{(2\pi)^3}\sqrt{\frac{M_N}{E_{p'}}}\:V(\Vec{p'},\Vec{p})\sqrt{\frac{M_N}{E_p}} 
\end{align}

and 
\begin{align}
   \widehat{T}(\Vec{p'},\Vec{p})\equiv\frac{1}{(2\pi)^3}\sqrt{\frac{M_N}{E_{p'}}}\:T(\Vec{p'},\Vec{p})\sqrt{\frac{M_N}{E_p}} \,, 
\end{align}
the BbS equation collapses into the usual, nonrelativistic Lippmann-Schwinger (LS) equation,
\begin{align}
  \widehat{T}(\Vec{p'},\Vec{p})&=\widehat{V}(\Vec{p'},\Vec{p})+\int{d^3\:p''}\:\widehat{V}(\Vec{p'},\Vec{p''})
    \:\frac{M_N}{p^2-p''^{2}+i\epsilon}\:\widehat{T}(\Vec{p''},\Vec{p}) \,.
    \label{eq_LS}
\end{align}

Iteration of $\widehat{V}$ in the LS equation, Eq.~(\ref{eq_LS}), requires cutting $\widehat{V}$ off for high momenta to avoid infinities. This is consistent with the fact that ChPT is a low-momentum expansion which is valid only for momenta $Q\leq\Lambda_b\approx 0.6$\:GeV. Therefore, the potential $\widehat{V}$ is multiplied with a regulator function $f(p',p)$,
\begin{equation}
    \widehat{V}(\Vec{p'},\Vec{p})\longmapsto \widehat{V}(\Vec{p'},\Vec{p})\:f(p',p) \,,
\end{equation}
with
\begin{equation}
    f(p',p)=exp[-(p'/\Lambda)^{2n}\:-(p/\Lambda)^{2n}] .
\end{equation}
For the chiral potentials applied in this investigation, we use
$\Lambda=500$ MeV~\cite{EMN17}.


\section{ Relevance of contact terms {\it versus} pion exchanges in lower partial waves}
\label{sec_relevance}

\begin{figure}\sidecaption
\includegraphics[height=12cm, width=8cm]{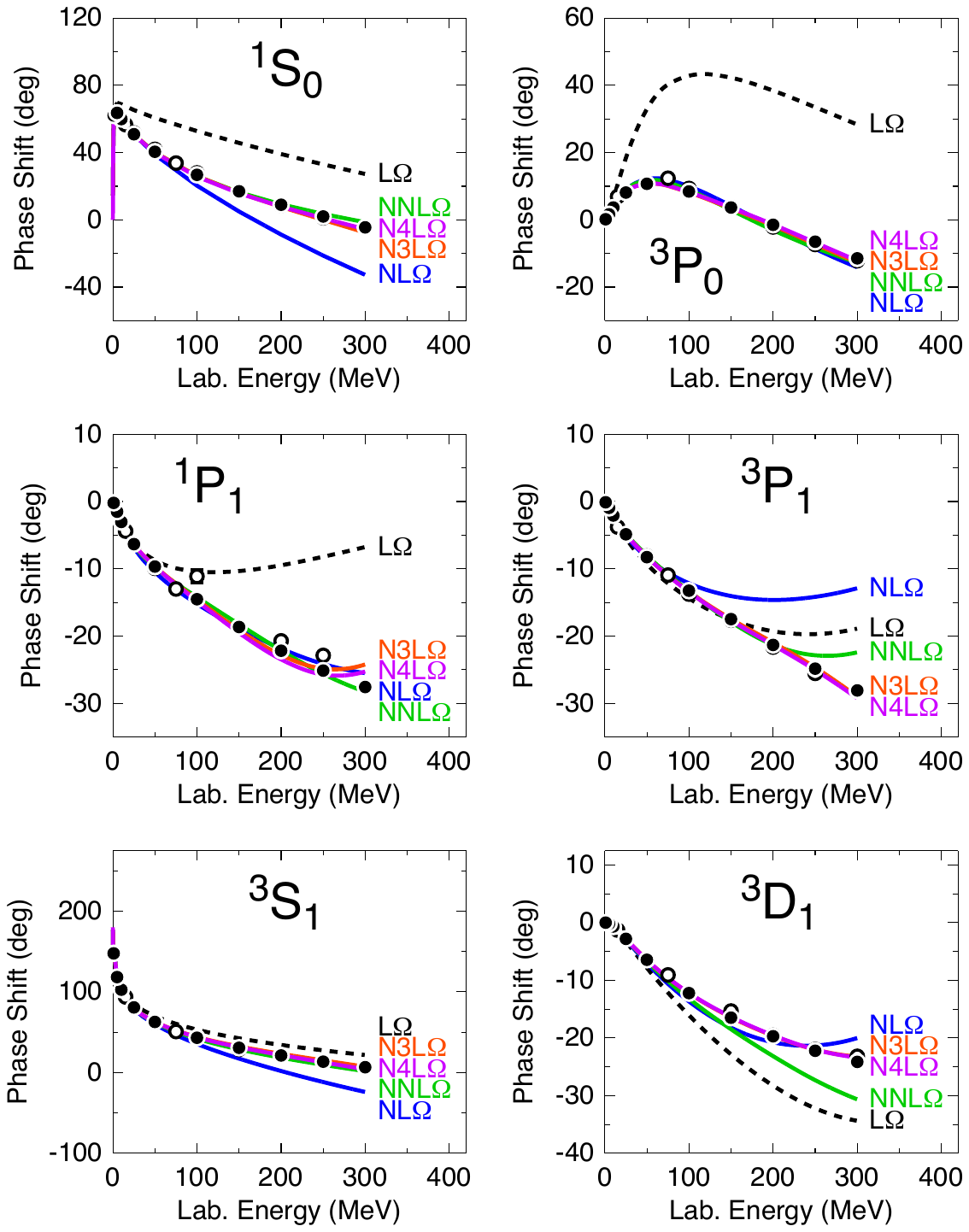}
\caption{Chiral expansion of $np$ scattering as represented by the phase shifts in  the
$J=0,1$ states.
 Five orders ranging from L$\Omega$ to N$^4$L$\Omega$ are shown as denoted. The solid dots and open circles are results from the Nijmegen multienergy $np$ phase shift analysis~\cite{Sto93} and GWU single-energy $np$ analysis~\cite{SP07}, respectively. (From Ref.~\cite{EMN17})}
\label{fig_ph1a}
\end{figure}

\begin{figure}\sidecaption
\vspace{-0.75cm}
\includegraphics[height=12 cm,width=8 cm]{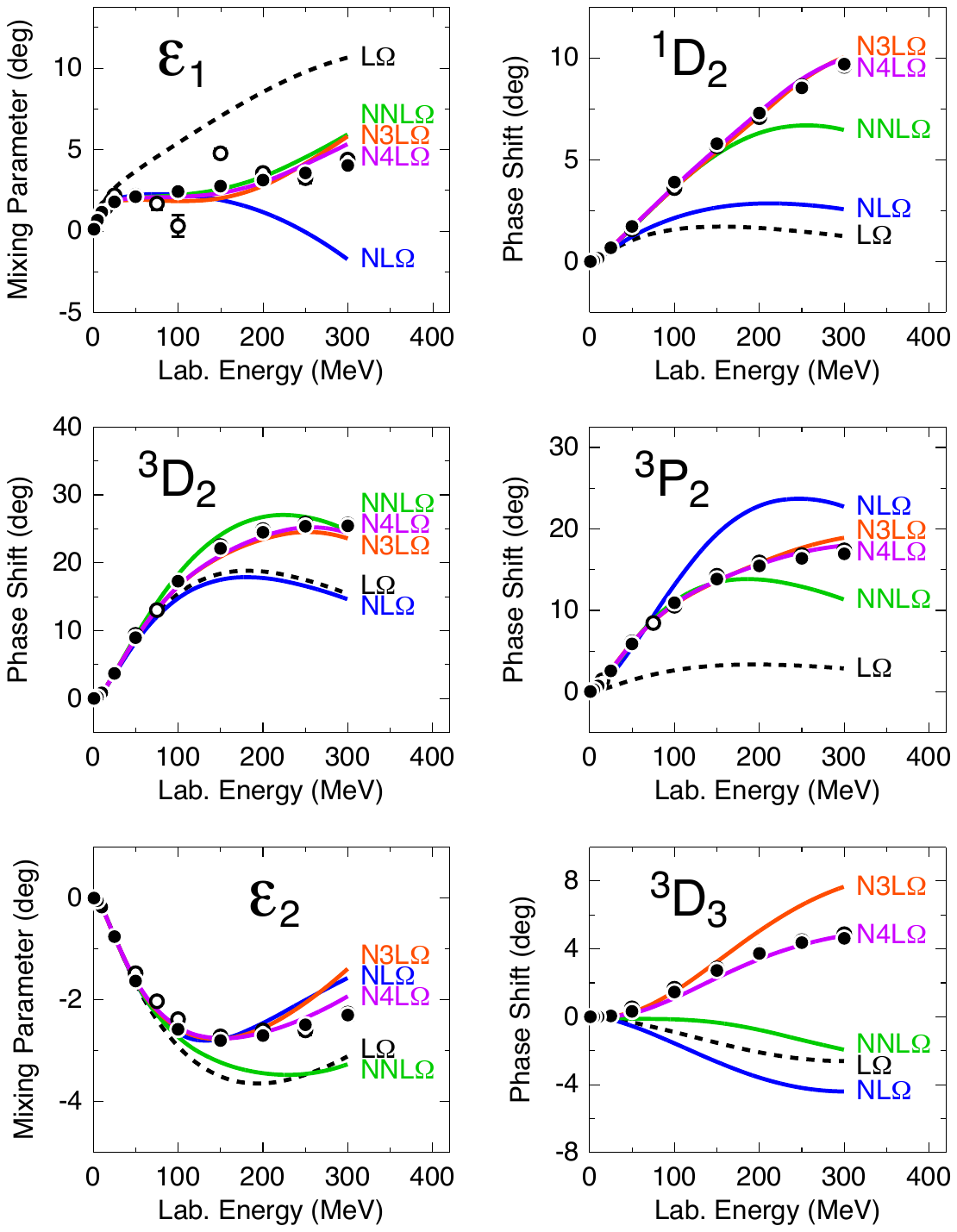}
\caption{Same as Fig.~\ref{fig_ph1a}, but for $J=2$, $\epsilon_1$, $\epsilon_2$, and
$^3D_3$.}
\label{fig_ph1b}
\end{figure}

In Ref.~\cite{EMN17}, $NN$ potentials through all orders from L$\Omega$ to N$^4$L$\Omega$ were constructed.
There are improvements in the reproduction of the empirical phase shifts as the orders increase and an excellent agreement is achieved
at orders N$^3$L$\Omega$ and N$^4$L$\Omega$, Figs.~\ref{fig_ph1a} and \ref{fig_ph1b}.
Similar results have been obtained by other 
researchers in the field~\cite{Pia15,Pia16,Car16,RKE18,Eks18,EKM15}.
Note that such fits involve 
two contacts at L$\Omega$ which contribute in $S$-waves,
nine contacts at NL$\Omega$ and NNL$\Omega$  which contribute up to $P$-waves, and
24 contacts at N$^3$L$\Omega$ and N$^4$L$\Omega$ which contribute up to $D$-waves
[cf.\ Eq.~(\ref{eq_ct})]. It is the purpose of this study  to analyze in detail the role of these contacts in the fits
of the $L\leq 2$ phase shifts up to N$^3$L$\Omega$.

As discussed,
the nuclear force consists essentially of two parts: the short range (Sect.~\ref{sec_short}) and the long range (Sect.~\ref{sec_long}).
In chiral EFT, the long-range is represented by one- and multi-pion exchanges, and
the short range is described by contact terms.
The lower partial waves are particularly sensitive to the short range and, in fact, at N$^3$L$\Omega$, four contact terms contribute to each $S$-wave, two to each $P$-wave, and one to each $D$-wave, \ Eq.~(\ref{eq_ct}).
There are no contact contributions in $F$ and higher partial waves---at N$^3$L$\Omega$.

As explained in the Introduction, since lower partial waves are sensitive to the short range potential, one may suspect that the contact contributions are dominant and simply override the pion-exchange contributions in lower partial waves which, on the other hand, are
the most important ones in applications of the potentials to nuclear structure and
reactions. 
 
 To shed light on this issue, we will now 
 systematically investigate the role of those contacts {\it versus} pion exchange in those lower partial waves of $NN$ scattering.

 In our $NN$ potential construction, the $\pi N$ LECs are not fit-parameters; they are held fixed at their values determined in $\pi N$ scattering, Table~\ref{tab_LEC_piN}. Therefore, the LECs of the $NN$ contacts are the only fit parameters available to optimize the reproduction of the $NN$ data (below 300 MeV laboratory energy). 
In this investigation, we will use the contact LECs to fit specific $NN$ low-energy parameters or phase shifts.
We will consider various scenarios, namely,
using contacts only or using contacts together with pion contributions of increasing chiral order. The failure to reproduce the $NN$ data by contacts only 
and the improvements that occur when (chiral) pion contributions are added will
reveal the relevance of chiral symmetry in those lower partial waves.
To obtain maximum insight into the role that contact terms can play,
 we will not follow here the rule that contact
and pion contributions should be of the same order. In fact, we may, for example,
consider contact contributions up to fourth order alone or with just the (lowest order) 1PE or low-order 2PE added,
to demonstrate what contacts can maximally achieve or not achieve.
For contacts and pion exchanges, we consider orders up to N$^3$L$\Omega$ (fourth order).

To keep it simple at the start, we begin with the partial waves that have only one contact, namely, $D$-waves and, then, proceed to the more elaborate cases, $P$ and $S$-waves.

 \subsection{$D$-waves}

To demonstrate the relevance of the pion exchange contributions ({\it versus} contacts)
 in $D$-waves, we consider the following cases for which we introduce
the short notation given in parenthesis.
\begin{itemize}
\item
Contact contribution only (ct1). 
\item  
L$\Omega$ pion exchange (i.~e., 1PE) only and no contact term (L0).
\item
L$\Omega$ 1PE plus contact term (L1).
\item
NL$\Omega$ pion exchanges only, no contact term (NL0).
\item
NL$\Omega$ pion exchanges plus contact term (NL1).
\item
NNL$\Omega$ pion exchanges only, no contact term (NNL0).
\item
NNL$\Omega$ pion exchanges plus contact term (NNL1).
\item
N$^3$L$\Omega$ pion exchanges plus contact term (NNNL1).
\end{itemize}
Our short notation (given in parenthesis) is designed such that the letters always indicate the order of the pion exchanges included and the integer states the number of contacts involved (from the contacts available
for the given partial wave).
Note that in $D$-waves, there is only one (fourth order) contact per partial wave available, Eq.~(\ref{eq_ct}). When we include this contact term, we fit it to the
empirical phase-shift at 50 MeV laboratory energy as determined in the Nijmegen phase-shift analysis~\cite{Sto93}. The values for the contact LECs so obtained are listed in 
Table~\ref{LEC_D-waves}. 

\begin{figure}\centering
\vspace{-2cm}
\subfloat{\includegraphics[width = 2.4 in]{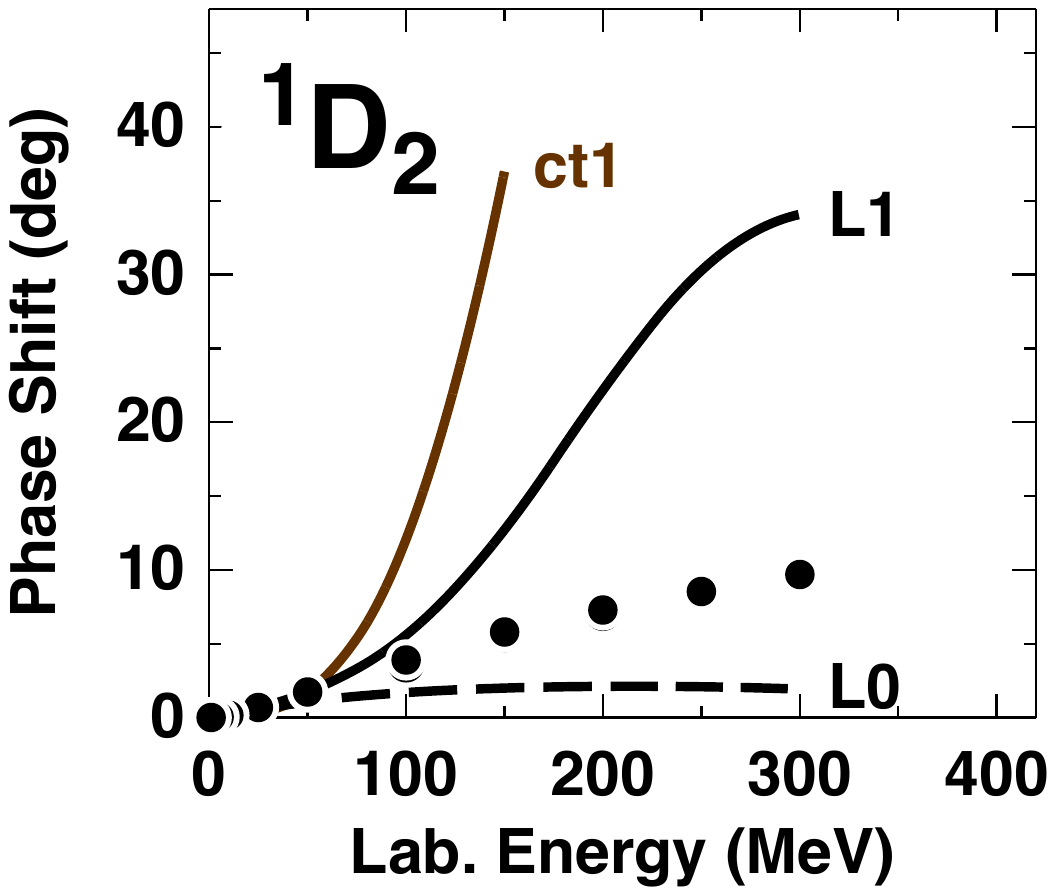}}
\subfloat{\includegraphics[width = 2.4 in]{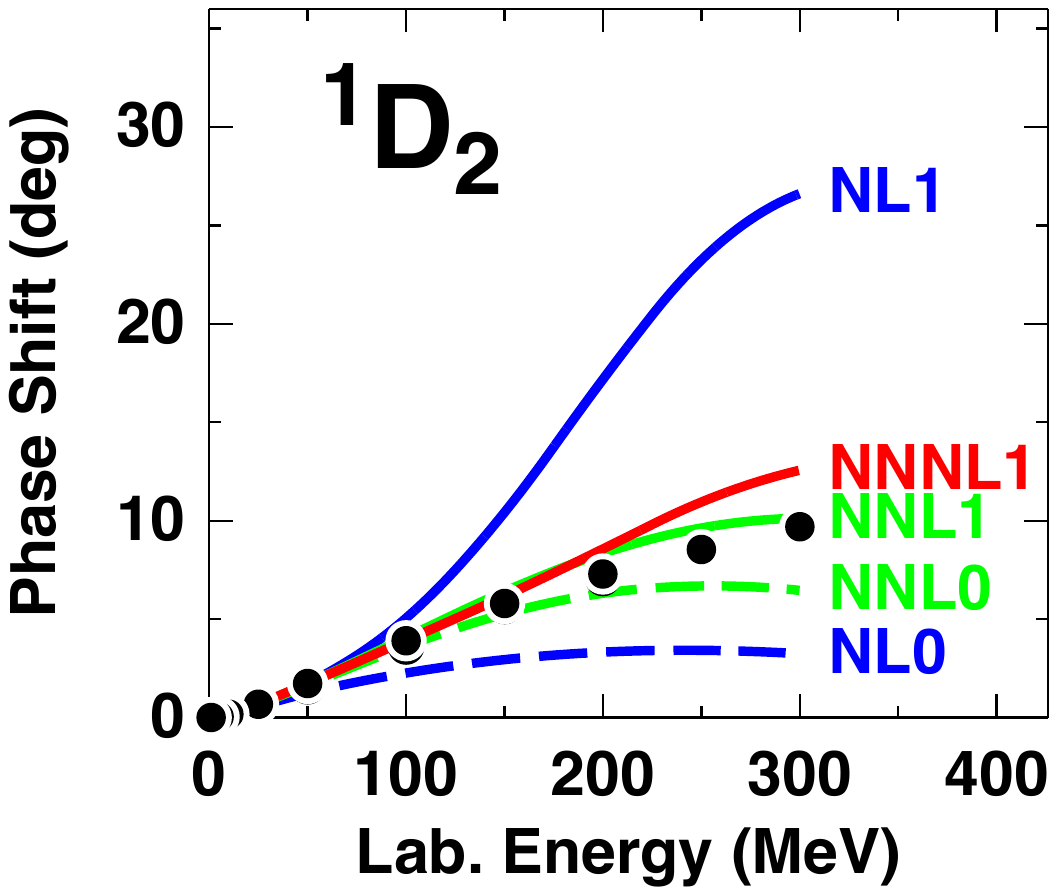}}
\vspace{-3.2cm}\\
\subfloat{\includegraphics[width = 2.4 in]{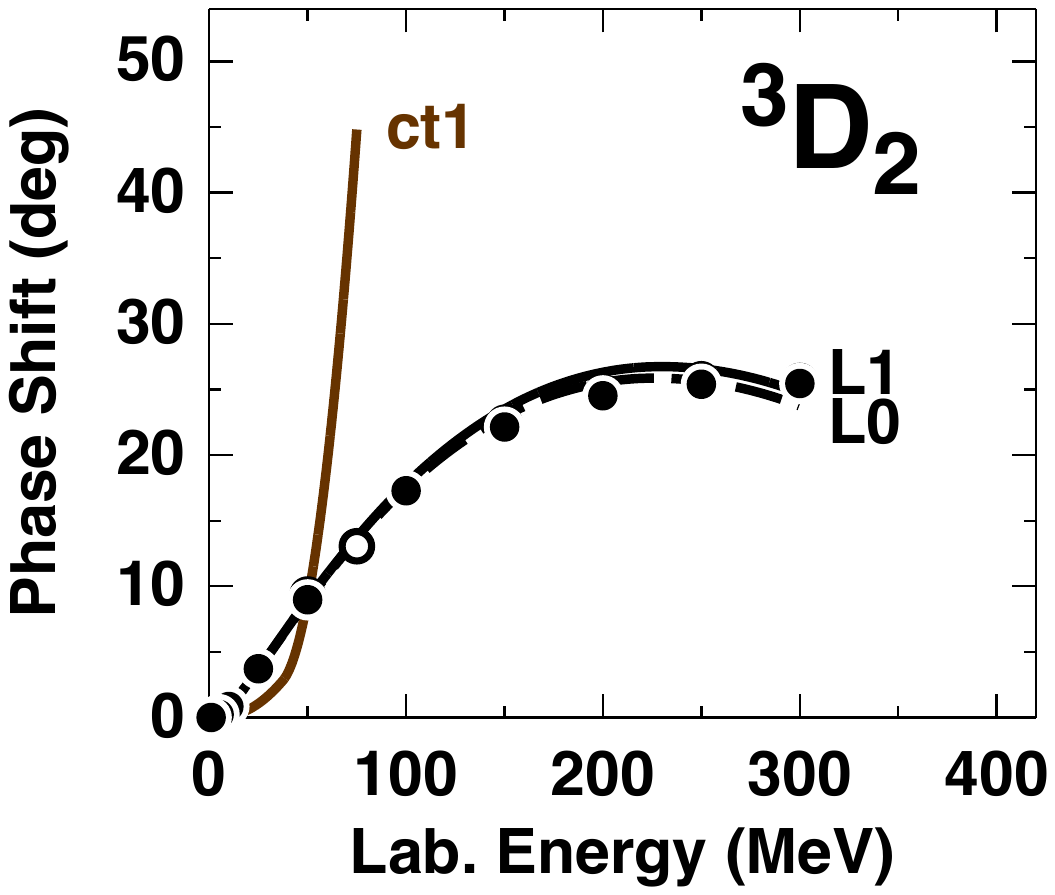}}
\subfloat{\includegraphics[width = 2.4 in]{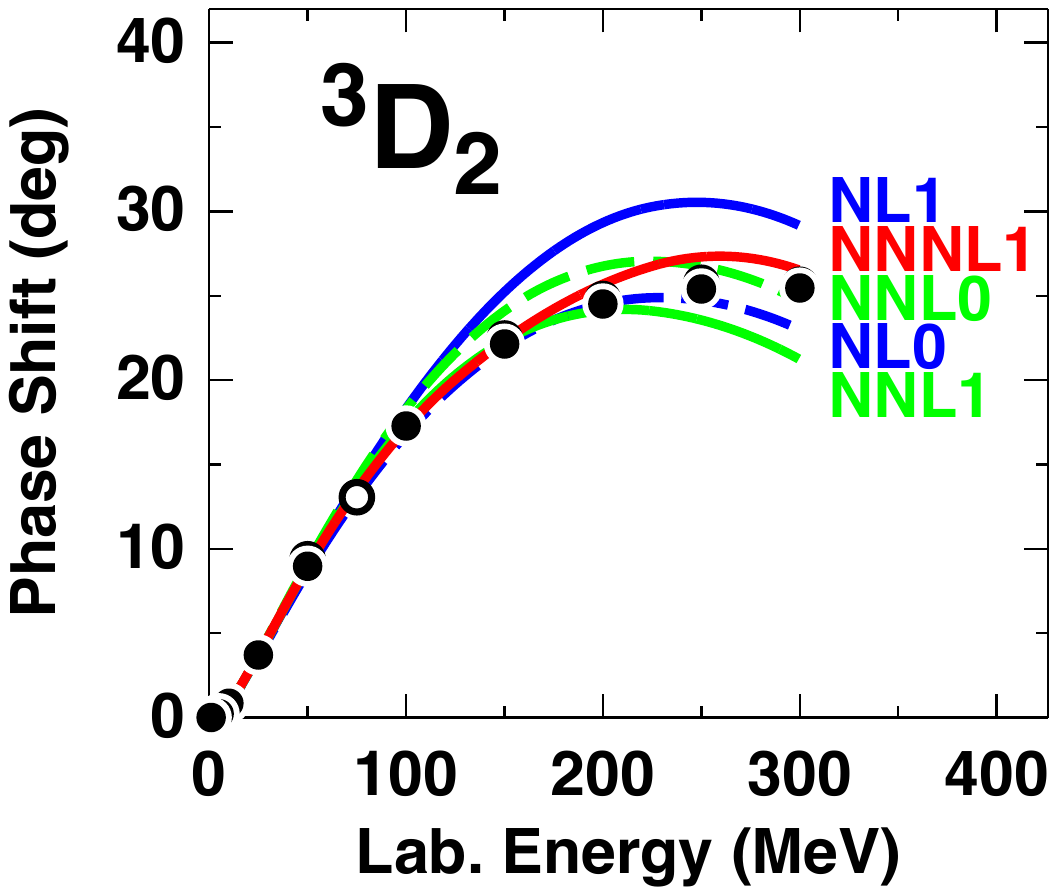}}
\vspace{-3.2cm}\\
\subfloat{\includegraphics[width = 2.4 in]{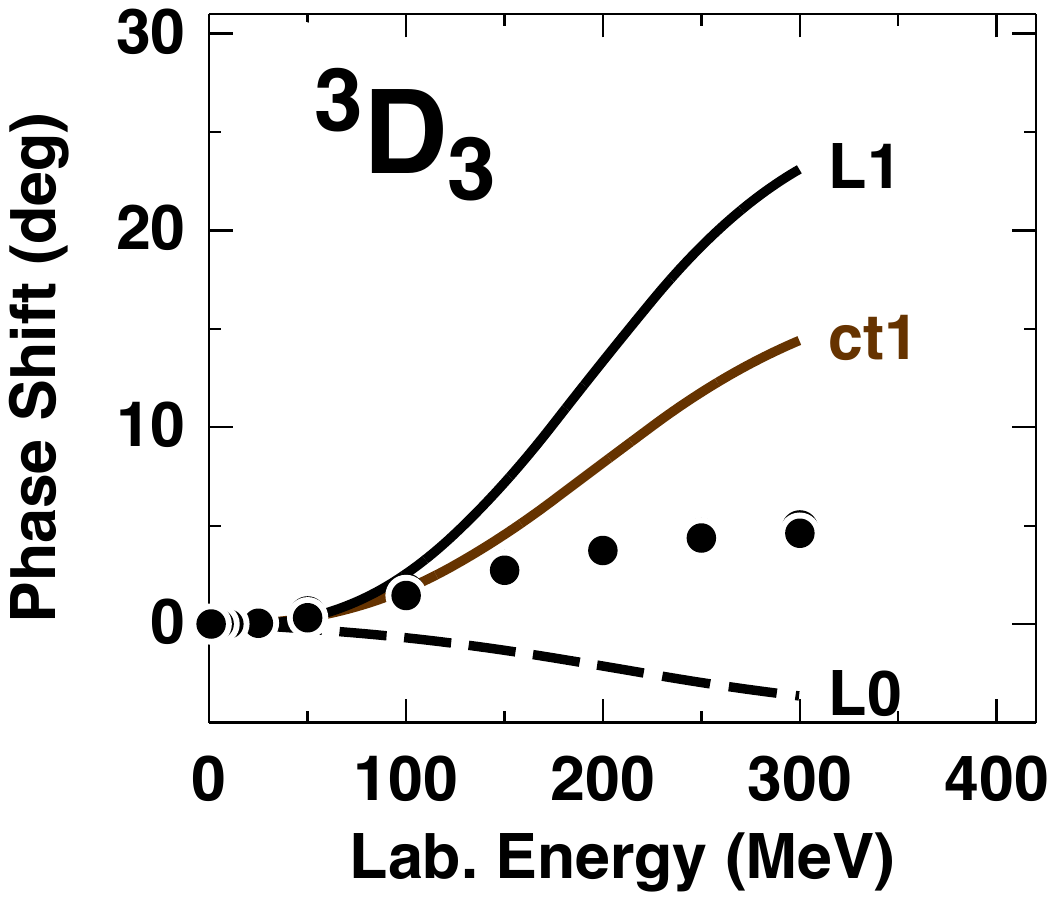}}
\subfloat{\includegraphics[width = 2.4 in]{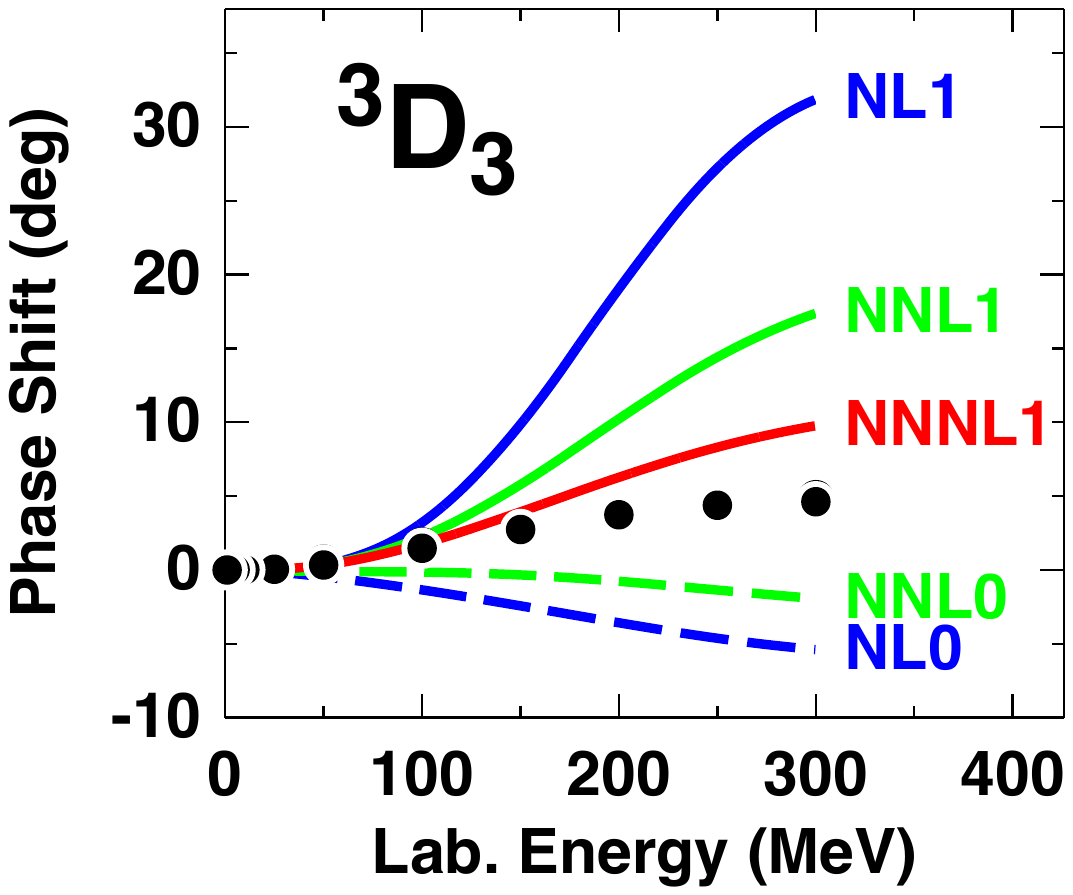}}
\vspace{-1.5 cm}\\
\caption{$D$-wave phase shifts of neutron-proton scattering 
for the various cases discussed in the text.
Solid dots and open circles as in Fig.~\ref{fig_ph1a}.}
\label{D-waves}
\end{figure}

Note that the chiral 2PE expressions at orders 
NL$\Omega$, Eqs.~(\ref{eq_2C}) and (\ref{eq_2T}),
and
NNL$\Omega$, Eqs.~(\ref{eq_3C}) and (\ref{eq_3T}),
include polynomial terms up to order $Q^3$~\cite{KBW97}, 
which do not contribute in $D$-waves [cf.\ Eq.~(\ref{eq_ct00}) and text below the equation].
Therefore, in the cases of L1, NL1, and NNL1, the $Q^4$ contacts are not renormalized and represent
the true corrections needed on top of the non-polynomial parts of the pion-exchanges, denoted
by L0, NL0, and NNL0, respectively (Table~\ref{LEC_D-waves} and Fig.~\ref{D-waves}).
 
  The situation is different at N$^3$L$\Omega$. 
  The subtracted dispersion integrals, Eq.~(\ref{eq_disp}), generate---besides the
non-polynomial parts---polynomial terms up to fourth order. 
Moreover, the other N$^3$L$\Omega$ 2PE contributions (Sect.~\ref{sec_n3lo})
and the contributions of Sect.~\ref{sec_rel}
  also include polynomial terms
of $\mathcal{O}(Q^4)$.
Thus, the fourth order contact term, we introduce to fit the phase shift at 50 MeV, includes
a compensation for the fourth order polynomial terms generated by the  2PE contributions. 
Therefore, in the case of NNNL1 of Table~\ref{LEC_D-waves}, the contact LECs are
 ``renormalized''. It is 
not just the correction needed besides the non-polynominal 2PE
contribution at N$^3$L$\Omega$ to fit the $D$-wave phase shifts at 50 MeV, and should not be interpreted that way. In fact, the large size of the NNNL1 contact LECs shown 
in Table~\ref{LEC_D-waves} indicate that the fourth order polynomial terms generated by
N$^3$L$\Omega$ pion contributions can be sizable. 

The phase shifts up to 300 MeV predicted for the various cases are
shown in Fig.~\ref{D-waves}.
Next we will discuss those phase shifts partial wave by partial wave.

\begin{table}\centering
    \caption{Contact LECs used for $D$-waves [cf.\ Eq.~(\ref{eq_ct})] in units of $10^4$ GeV$^{-6}$.}
\begin{tabular*}{\textwidth}{@{\extracolsep{\fill}}llll}
    \hline \hline
    \noalign{\smallskip}
     case &  $D_{^{1}D_2}$ &  $D_{^{3}D_2}$ & $D_{^{3}D_3}$ \\
         \noalign{\smallskip}
      \hline
          \noalign{\smallskip}
        ct1 & -3.2575 & -5.7202 & -1.0130 \\
            \noalign{\smallskip}
        L1 & -1.6165 & -0.0578 & -1.7843 \\
             \noalign{\smallskip}
        NL1 & -1.2045 & -0.3464 & -2.3773 \\
             \noalign{\smallskip}
        NNL1 & -0.2068 & 0.2023 & -1.3345 \\
                   \noalign{\smallskip}
        NNNL1 & -2.088 & -3.3804 & -1.4764 \\
         \noalign{\smallskip}
      \hline \hline
\end{tabular*}
    \label{LEC_D-waves}
\end{table}

\subsubsection{The $^1D_2$-wave}
We start with the left $^1D_2$ frame in Fig.~\ref{D-waves}. When only the contact term is applied
and no pion-exchanges (curve ct1) then the phase shift increases dramatically with energy indicating
that the contact contribution is of very short range and completely inadequate to describe this $D$-wave.
1PE is weak (curve L0). Adding the contact to 1PE brings the phase shift up, but too much since obviously the contact is dominant. When 2PE contributions are added (right $^1D_2$ frame),
the description improves with increasing order. While the NL$\Omega$ 2PE is weak and, therefore, does not lead to much improvement (cf.\ NL0 and NL1), the NNL$\Omega$ 2PE is known to provide a realistic
intermediate-range attraction and together
with the contact leads to a quantitative description (curve NNL1),
and so does NNNL1. The conclusion is that the contact alone can by no means describe $^1D_2$.
The strong intermediate-range attraction provided by chiral 2PE at NNL$\Omega$ and N$^3$L$\Omega$ is crucial.
As the small contact LEC in the case of NNL1 reveals (Table~\ref{LEC_D-waves}), the contact
contribution is minor, while chiral 2PE rules. This example demonstrates that even when
a contact term is involved, chirality can still be the major factor and show a clear signature.

\subsubsection{The $^3D_2$-wave}
Also in the $^3D_2$-wave, the contact contribution alone 
(cf.\ ct1 curve in the left $^3D_2$ frame in Fig.~\ref{D-waves})
leads to a dramatically wrong description.
In this particular partial wave, the 1PE (L0 curve) happens to play a dominant role,
because the matrix element of the tensor operator is 2 in this state which, in addition, is mutiplied by (-3) from the $ \bm{\tau}_1 \cdot \bm{\tau}_2 $ factor, resulting in an overall factor of (-6) for the pion tensor potential. As the L0 curve reveals, this large tensor contribution alone, essentially, explains the $^3D_2$-wave. 2PE contributions play only a minor role (cf.\ right $^3D_2$ frame),
because the (mainly) central forces provided by 2PE are small as compared to the huge tensor force contribution from 1PE  in this particuler wave.
This scenario leaves little room for contact contributions.
One-pion-exchange, the most pronounced expression of chiral symmetry, rules this wave.

\subsubsection{The $^3D_3$-wave}
The cases ct1, L0, and L1 are inadequate 
similarly to what we have seen in $^1D_2$.
The 2PE contributions at NL$\Omega$ and NNL$\Omega$ without and with contact contribution 
(NL0, NNL0 and NL1, NNL1, respectively)
do not lead to much improvement. Finally, with NNNL1 a more realistic result starts to develop. A quantitative
description has to wait for N$^4$L$\Omega$ as demonstrated in the $^3D_3$ frame of Fig.~\ref{fig_ph1b}. 
In any case, the contact alone cannot describe the $^3D_3$ wave, since the contact contribution
 is too short-ranged.
Substantial intermediate-range attraction is needed, which only chiral 2PE can provide.

\subsubsection{The $^3D_1$-wave}
Since the $^3D_1$ wave is coupled to $^3S_1$, we will discuss it in conjunction with the coupled $^3S_1$-$^3D_1$-$\epsilon_1$ system, below.

\subsubsection{$D$-waves summary}
Contacts alone can not reproduce $D$-waves (cf.\ all the ct1 cases in the left column of Fig.~\ref{D-waves}), because of the short-range nature of contact contributions, which are ill-suited for $D$-waves.
The strong intermediate-range attraction provided by chiral 2PE at NNL$\Omega$ and N$^3$L$\Omega$ is crucial, unless the 1PE tensor force is dominant, which also is
a reflection of chiral symmetry.
For exact fits, contact corrections are needed, but they are very small.
Thus, in spite of contributions from contacts, chirality makes the largest imprint on $D$-waves.

The $D$-waves are, in fact, an interesting case.
On the one hand, they are not so peripheral that the (very long-ranged) 1PE is dominant and,
on the other hand, their orbital angular momentum is large enough to prevent them
from being too sensitive to the (short-ranged) contact potential.
Thus, the $D$-waves are a true window on the intermediate range.
Consequently, they test the reality of the (internediate-ranged) 2PE as produced by chiral symmetry.
In particular, the  $^1D_2$-wave demonstrates that this test is passed well.

\subsection{$P$-waves}

In $P$-waves, we have two contacts available per partial wave; one is of order two, $C_\alpha$, and the other one is of order four, $D_\alpha$ [cf.\ Eq.~(\ref{eq_ct})].  We then consider the following cases with
the short notation given in parenthesis.
\begin{itemize}
\item
One contact contribution and nothing else (ct1). 
\item
Two contact contributions (ct2). 
\item  
L$\Omega$ pion exchange (i.~e., 1PE) only and no contact term (L0).
\item
L$\Omega$ 1PE plus one contact term (L1).
\item
L$\Omega$ 1PE plus two contact terms (L2).
\item
NL$\Omega$ pion exchanges plus one contact term (NL1).
\item
NL$\Omega$ pion exchanges plus two contact terms (NL2).
\item
NNL$\Omega$ pion exchanges plus one contact term (NNL1).
\item
NNL$\Omega$ pion exchanges plus two contact terms (NNL2).
\item
N$^3$L$\Omega$ pion exchanges plus two contact terms (NNNL2).
\end{itemize}

The values for the contact LECs used in the various cases are listed in 
Table~\ref{LEC_P-waves}. 

As mentioned, the chiral 2PE expressions at orders NL$\Omega$ and NNL$\Omega$
include polynomial terms of order $Q^2$ and the 2PE expressions at order N$^3$L$\Omega$
include polynomial terms up to order $Q^4$.
$\mathcal{O}(Q^2)$ and $\mathcal{O}(Q^4)$ polynomial terms do not vanish in $P$-waves [Eq.~(\ref{eq_ct00})].
These terms are absorbed by the 
second and fourth order contact terms. Therefore, the minimal number of contacts to be applied at
 NL$\Omega$ and NNL$\Omega$ is one (second order) contact and two (second and fourth order) contacts at N$^3$L$\Omega$.
Thus, the contact LECs shown in Table~\ref{LEC_P-waves} for  NL1, NNL1, and NNNL2
are not just the corrections needed besides the genuine 2PE
contributions and their size does not 
reflect the size of ``what is missing''.
However, in the cases NL2 and NNL2, the second contact included, $D_\alpha$ (fourth order contact), is not renormalized (since NL$\Omega$ and NNL$\Omega$ 2PE does not generate
$Q^4$ polynomials)  and, therfore, reflects a true fourth order correction.

\begin{figure}\centering
\vspace{-2.0cm}
\subfloat{\includegraphics[width = 2.4 in]{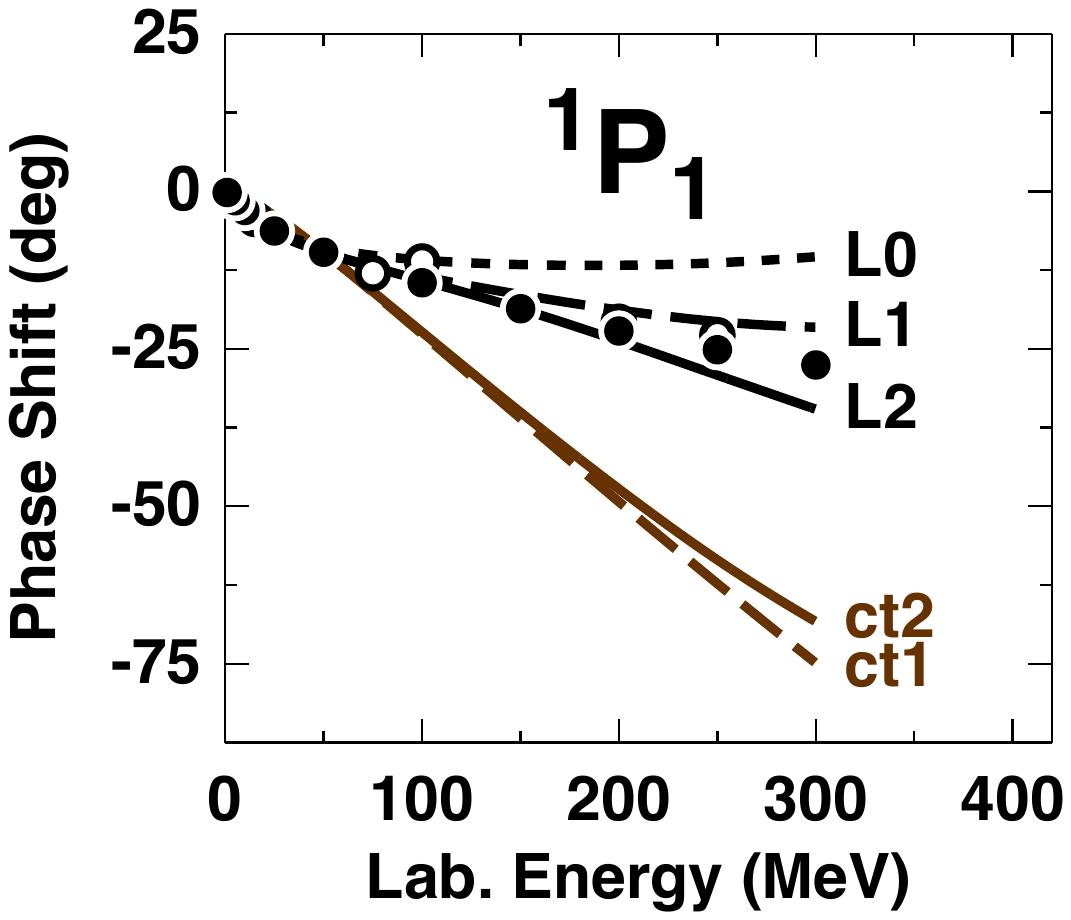}}
\subfloat{\includegraphics[width = 2.4 in]{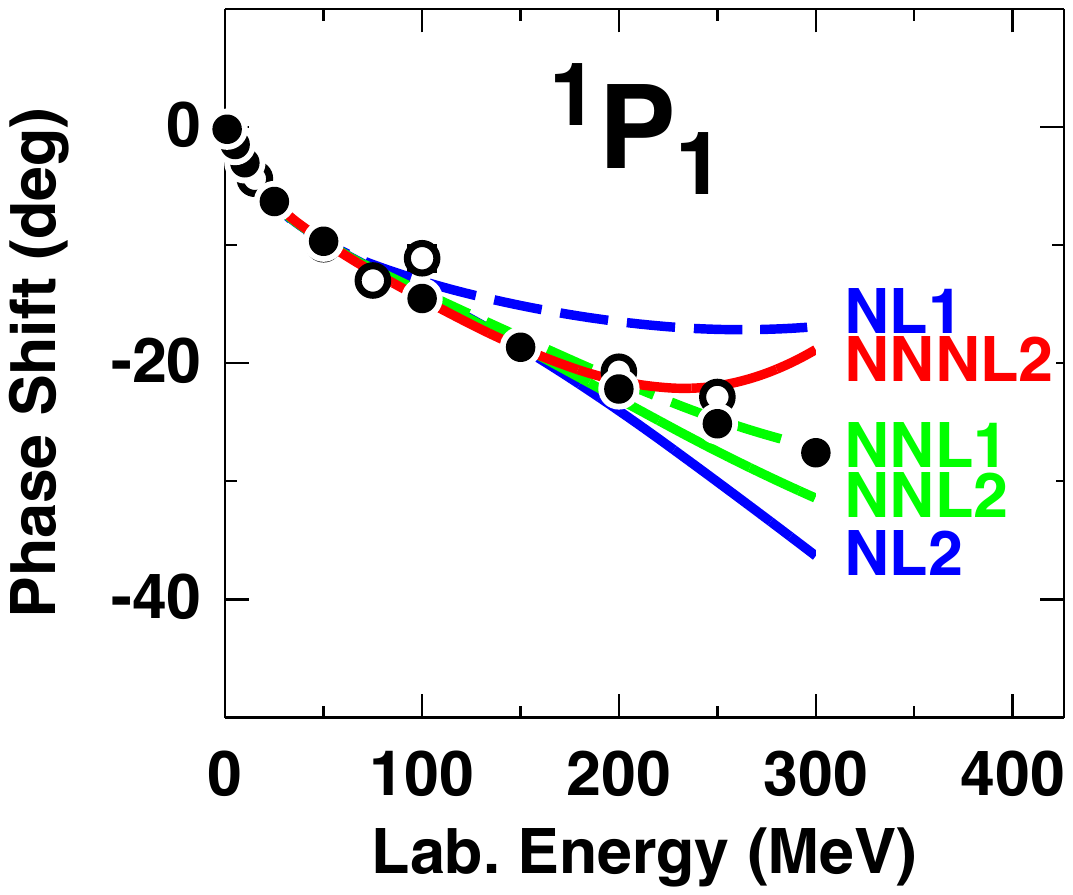}}
\vspace{-3.2cm}\\
\subfloat{\includegraphics[width = 2.4 in]{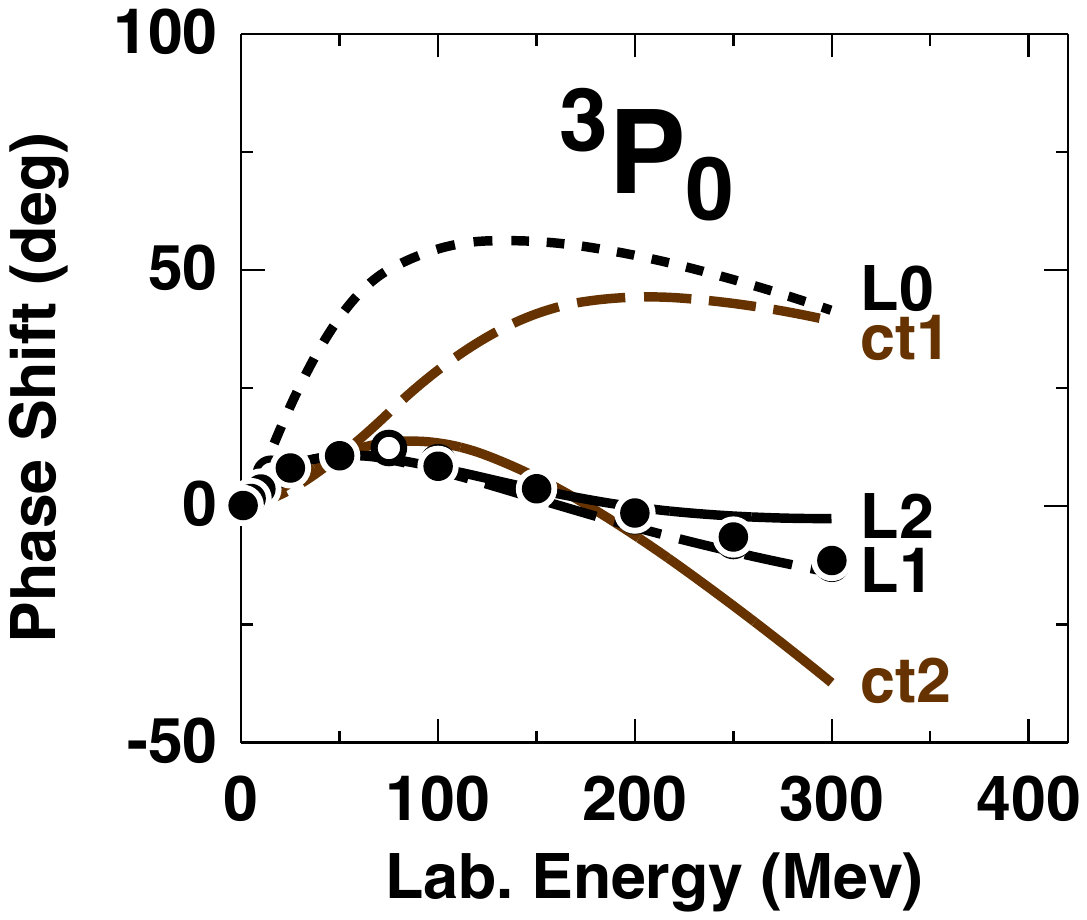}}
\subfloat{\includegraphics[width = 2.4 in]{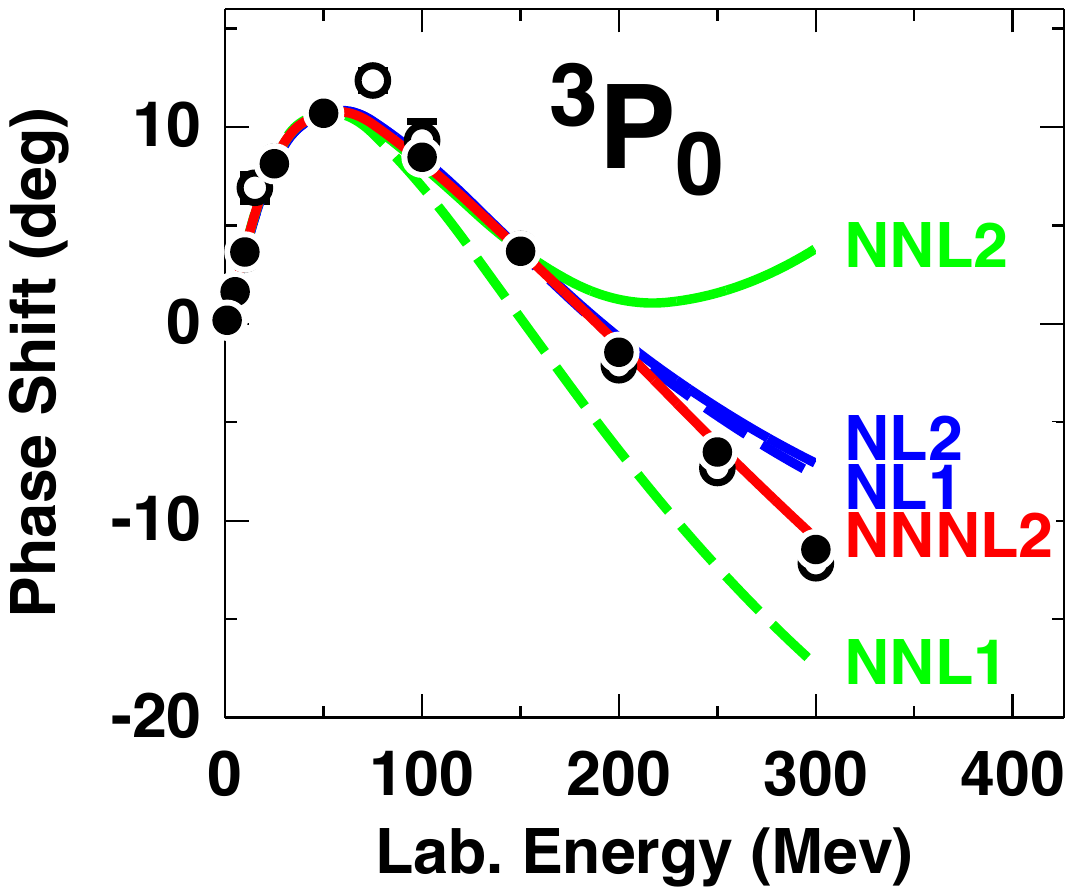}}
\vspace{-3.2cm}\\
\subfloat{\includegraphics[width = 2.4 in]{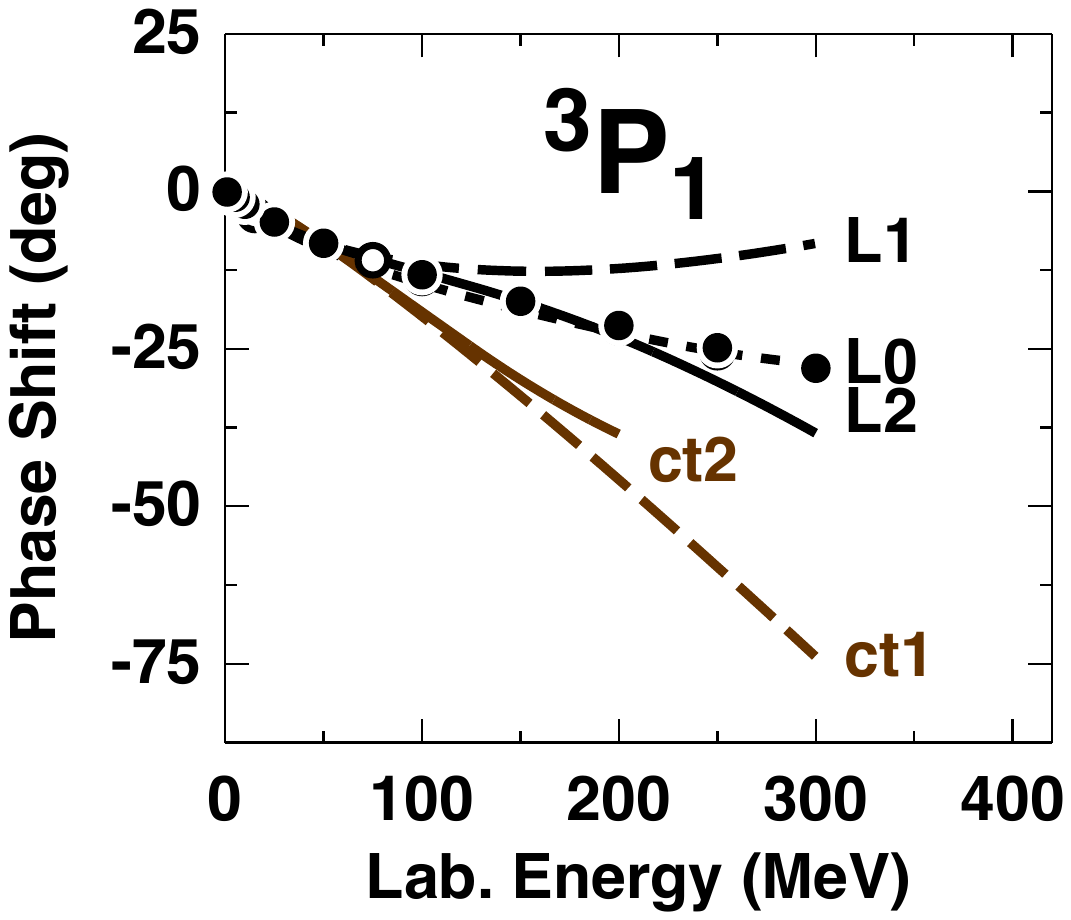}}
\subfloat{\includegraphics[width = 2.4 in]{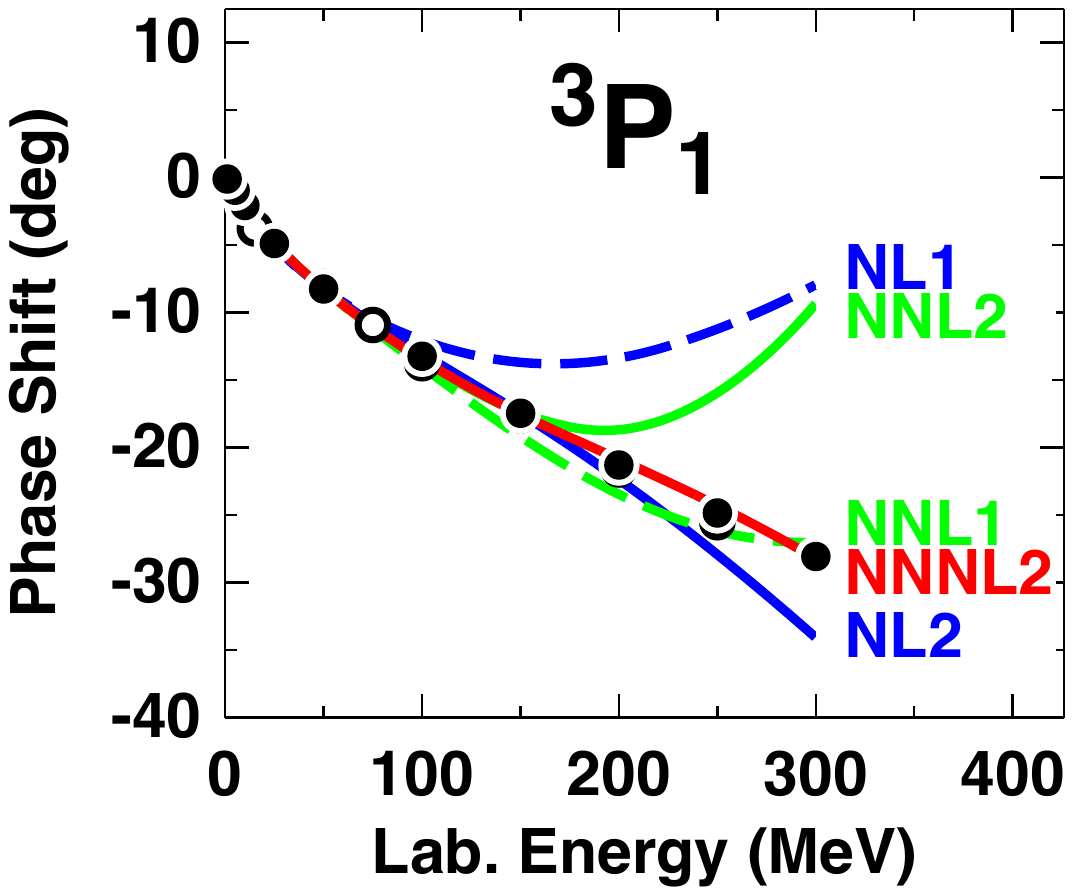}}
\vspace{-3.2 cm}\\
\subfloat{\includegraphics[width = 2.4 in]{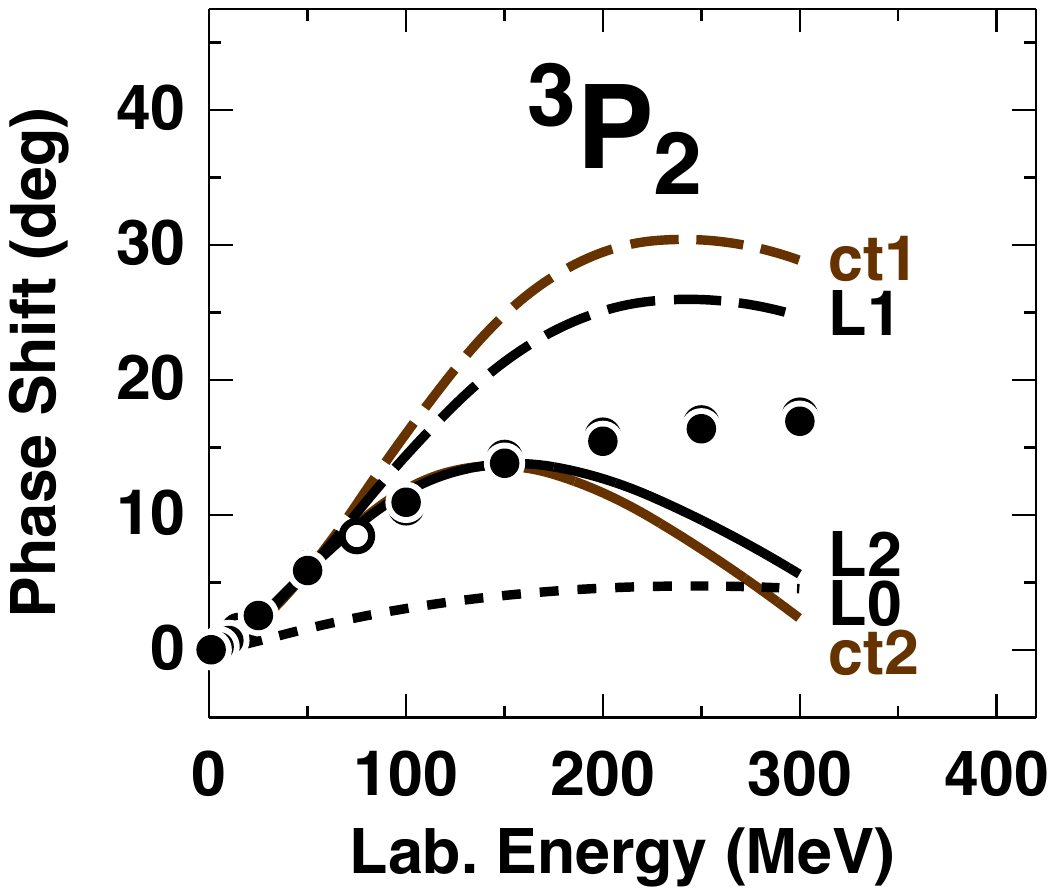}}
\subfloat{\includegraphics[width = 2.4 in]{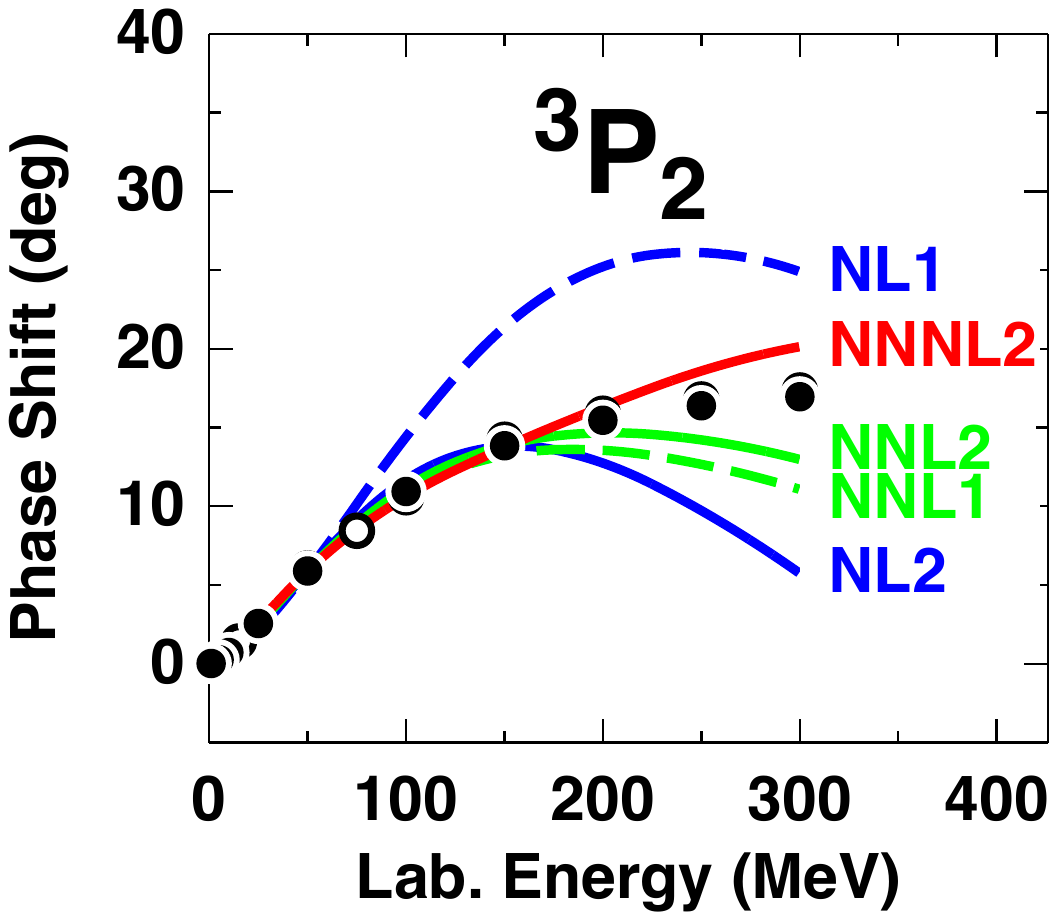}}
\vspace{-1.5 cm}\\
\caption{$P$-wave phase shifts of neutron-proton scattering 
for the various cases discussed in the text.
Solid dots and open circles as in Fig.~\ref{fig_ph1a}.}
\label{P-waves}
\end{figure}

The phase shifts up to 300 MeV that result from the various $P$-wave cases are
shown in Fig.~\ref{P-waves}, which we will discuss now.

\begin{table}\centering
    \caption{Contact LECs used in $P$-waves [cf.\ Eq.~(\ref{eq_ct})].
     Second order contacts, $C_\alpha$, are in units of 10$^4$ GeV$^{-4}$, while
      fourth order contacts, $D_\alpha$, 
    are in units of 10$^4$ GeV$^{-6}$.}
\begin{tabular*}{\textwidth}{@{\extracolsep{\fill}}lllllllll}
    \hline \hline
     \noalign{\smallskip}
             & \multicolumn{2}{c}{$^1P_1$}    & \multicolumn{2}{c}{$^3P_0$}  
               & \multicolumn{2}{c}{$^3P_1$}    & \multicolumn{2}{c}{$^3P_2$}
               \\
                \noalign{\smallskip}
               \cline{2-3}        \cline{4-5}        \cline{6-7}        \cline{8-9}   
                \noalign{\smallskip}
     case & $C_{^{1}P_1}$ & $D_{^{1}P_1}$&  $C_{^{3}P_0}$ &  $D_{^{3}P_0}$ & $C_{^{3}P_1}$ 
      & $D_{^{3}P_1}$ & $C_{^{3}P_2}$  & $D_{^{3}P_2}$
      \\
      \noalign{\smallskip}
      \hline \hline
       \noalign{\smallskip}
        ct1 & 6.5533 & 0 & -0.4631 & 0 & 4.3248 & 0 & -0.3256 & 0 \\
         \noalign{\smallskip}
        ct2 &2.17 & -5.0 & -0.874 & 10.0 & 1.4127& -5.0 & -0.4766 & 1.6 \\
            \noalign{\smallskip}
        L1 & 0.1349 & 0 & 0.8463 & 0 & -0.1732 & 0 &-0.2302 & 0 \\
            \noalign{\smallskip}
         L2&0.1613 & 0.95 &0.8531 & -0.55 &-0.1480 & 1.58 &-0.3300& 1.1 \\
             \noalign{\smallskip}
        NL1 & 0.2295& 0 & 1.3228 & 0 & -0.4607 & 0 &-0.2203 & 0 \\
            \noalign{\smallskip}
        NL2 &0.2664 & 1.45 &1.3234 & -0.03 &-0.4352 & 1.2 &-0.3203 & 1.1 \\
            \noalign{\smallskip}
        NNL1 &0.1821 & 0 & 1.1415 & 0 & -0.7851 & 0 &-0.6333 & 0 \\
            \noalign{\smallskip}
        NNL2 &0.1912 & 0.3 &1.1495 & -0.95 &-0.8133 & -0.58 &-0.6251 & -0.1 \\
            \noalign{\smallskip}
        NNNL2 & 0.1933& 9.72 & 1.1883 & 4.92 & -0.8105 & 4.74 &-0.7464 & 5.95 \\
            \noalign{\smallskip}
      \hline \hline
\end{tabular*}
    \label{LEC_P-waves}
\end{table}

When we apply only one contact, we use the order two one and fit it to the
empirical phase-shift at 50 MeV laboratory energy as determined in the Nijmegen phase-shift analysis~\cite{Sto93}. When both contacts are involved, we fit the empirical phase-shifts at 50 MeV and 150 MeV (if possible).

Obviously, with just one contact term and no pion contributions (cases ct1 of the left column of Fig.~\ref{P-waves}) the description 
is grossly wrong in all $P$-waves. Adding the second contact does not lead to any improvement in 
$^1P_1$ and $^3P_1$ and, in fact, in these two cases it is not possible to fit the phase shift at 150 MeV. The $^3P_0$ and $^3P_2$ partial waves improve with the second contact, but are not any close to a quantitative description. 
Adding 1PE (L0) together with one or two contacts (L1, L2) brings about considerable improvement in most $P$-waves. 
Turning to the frames of the right column of Fig.~\ref{P-waves} where the 2PE exchanges of various orders are added,
we observe order by order improvement. $^1P_1$ is described well in the cases of NNL1 and NNL2, while the other partial waves assume a quantitative character only
when the powerfull 2PE at N$^3$L$\Omega$ is added (case NNNL2). 

In summary, contacts alone are inadequate to describe $P$-waves. 1PE brings improvement, 
but strong chiral 2PE is needed for a quantitative description of $P$-waves.
Thus, a clear signature of chiral symmetry can be identified in $P$-waves.

A note is in place on $^3P_2$, since it is coupled with $^3F_2$ and $\epsilon_2$ through
 the contact LEC  $D_{^3P_2-^3F_2}$, Eq.~(\ref{eq_ct}). We found that the latter parameter
has only a weak effect on the $^3P_2$ phase shift and, therefore, 
we decided to leave it out of our considerations.
We kept it at zero.

\subsection{The $^1S_0$-wave}

\begin{table}\centering
 \caption{Columns two to five show the contact LECs used in the ${{}^{1}S_0}$ wave [cf.\ Eq.~(\ref{eq_ct})].
 The zeroth order contact $\widetilde{C}_{{}^{1}S_0}$ is in units of $10^4$ GeV$^{-2}$;
the second order contact $C_{{}^{1}S_0}$ in units of $10^4$ GeV$^{-4}$; 
and fourth order contacts $\widehat{D}_{{}^{1}S_0}$ and $D_{{}^{1}S_0}$
 in units of $10^4$ GeV$^{-6}$.
 Column six and seven display the $np$ scattering length, $a_{np}$, and effective range, $r_{np}$,
 in the ${{}^{1}S_0}$ state.  }
\begin{tabular*}{\textwidth}{@{\extracolsep{\fill}}lllllll}
    \hline \hline
    \noalign{\smallskip}
     case & $\widetilde{C}_{{}^{1}S_0}$ &  $C_{{}^{1}S_0}$ & $\widehat{D}_{{}^{1}S_0}$ & $D_{{}^{1}S_0}$ & $a_{np}$ (fm) & $r_{np}$ (fm) 
\smallskip
     \\
     \hline
      \noalign{\smallskip}
        ct1 &-0.063985& 0 &0 & 0 & -23.74 & 0.69\\
         \noalign{\smallskip}
        ct2 &0.475799& 4.0 & 0& 0 & -23.74 & 2.37 \\
         \noalign{\smallskip}
        ct3 &-0.158301& 2.0 &-6.0 & 0 & -23.74 & 2.66 \\
         \noalign{\smallskip}
        L1 & -0.109340& 0& 0&0 & -23.74 & 1.73  \\
         \noalign{\smallskip}
        L2&-0.130919&1.33&0&0 & -23.74 & 2.70\\
         \noalign{\smallskip}
         NL2 &-0.146214 &1.815&0&0 & -23.74 & 2.70 \\
          \noalign{\smallskip}
        NNL2 &-0.152032&2.36&0&0 & -23.74 & 2.70 \\
         \noalign{\smallskip}
        NNNL4 &-0.139563&2.417&-2.332&-16.74 & -23.74 & 2.70 \\
        \noalign{\smallskip}
      \hline \hline
\end{tabular*}
    \label{LEC_1s0-wave}
\end{table}

\begin{figure}\centering
\vspace{-1.0cm}
\subfloat{\includegraphics[width = 2.4in]{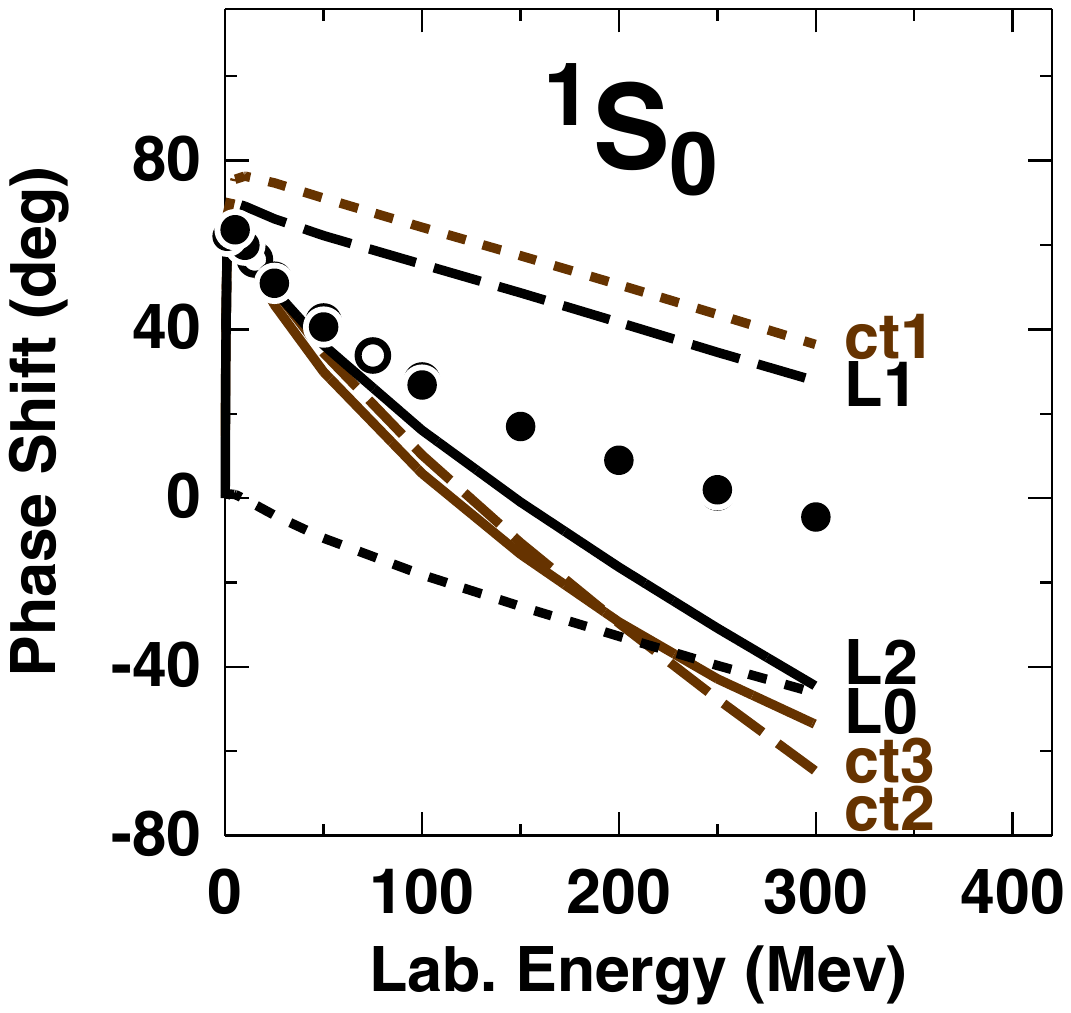}}
\subfloat{\includegraphics[width = 2.4 in]{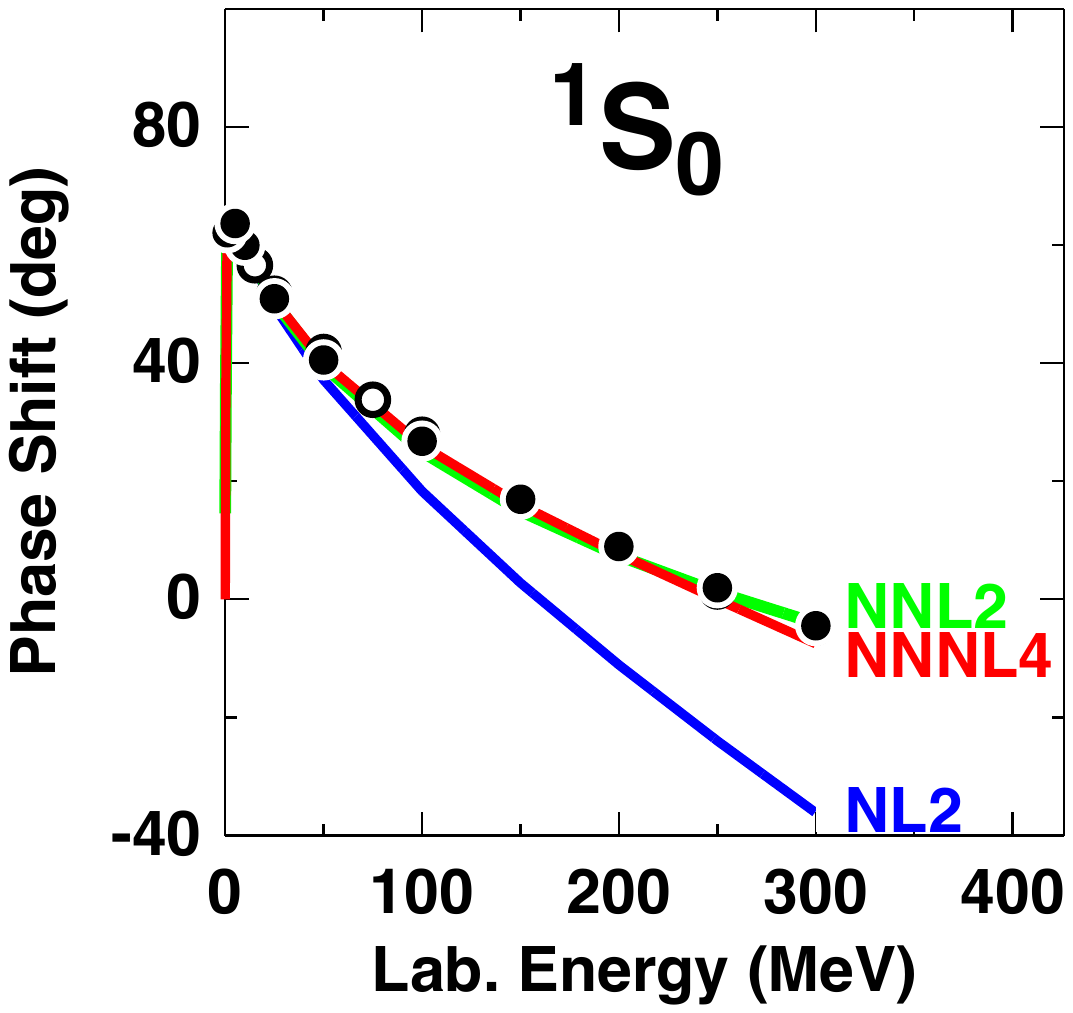}}
\vspace{-1.7cm}\\
\caption{$^1S_0$ phase shifts of neutron-proton scattering 
for the various cases discussed in the text.
Solid dots and open circles as in Fig.~\ref{fig_ph1a}.}
\label{1s0-wave}
\end{figure}

In the $^1S_0$ wave, we have available a total of four contact terms [cf.\ Eq.~(\ref{eq_ct})], namely, 
 one zeroth order contact, $\widetilde{C}_{{}^{1}S_0}$,
one second order contact, $C_{{}^{1}S_0}$, and
two fourth order contacts, $\widehat{D}_{{}^{1}S_0}$ and $D_{{}^{1}S_0}$.
When we use only one contact, we pick the zeroth order one and fit it to the $np$
$^1S_0$ scattering length, $a_{np}=-23.74$ MeV. When we apply two contacts, we fit, besides the scattering length, the $^1S_0$ $np$ effective range parameter, 
$r_{np}=2.70\pm 0.05$ MeV. With three parameters, we also try to reproduce (if possible) the
empirical phase-shift at 50 MeV laboratory energy as determined in the Nijmegen phase-shift analysis~\cite{Sto93} and, with four parameters, the phase shift at
150 MeV is included in the fit. 
We consider the following cases with
the short notation given in parenthesis.
\begin{itemize}
\item
One contact contribution, and nothing else (ct1). 
\item
Two contact contributions (ct2). 
\item
Three contact contributions (ct3). 
\item
L$\Omega$ 1PE and no contact term (L0).
\item
L$\Omega$ 1PE plus one contact term (L1).
\item
L$\Omega$ 1PE plus two contact terms (L2).
\item
NL$\Omega$ pion exchanges plus two contact terms (NL2).
\item
NNL$\Omega$ pion exchanges plus two contact terms (NNL2).
\item
N$^3$L$\Omega$ pion exchanges plus four contact terms (NNNL4).
\end{itemize}

The values for the contact LECs are listed in 
Table~\ref{LEC_1s0-wave} and
the phase shifts up to 300 MeV that result from the various $^1S_0$ cases are
shown in Fig.~\ref{1s0-wave}, which we will discuss now.

When only one contact term is used (fit to $a_{np}$) and no pion contributions (case ct1), then the 
$^1S_0$ phase shifts for intermediate energies are far above the data. Adding more contacts (cases ct2 and ct3) moves those predictions below the data. The prediction with four contacts is essentially the same as with three contacts and, therefore, not shown. Clearly, contacts alone cannot describe the $^1S_0$ wave at  intermediate energies, no matter how many contacts one is using. 1PE alone (L0) is small
and adding to it one or two contacts (cases L1 and  L2) brings about predictions
that are very similar to the coresponding cases with contacts alone (ct1 and ct2) and, again, adding more contacts does essentially not change anything.
Thus, in $^1S_0$, 1PE is obviously of very limited relevance, except for the effective range parameter, $r_{np}$, which is improved by 1PE (cf.\ Table~\ref{LEC_1s0-wave}).
The strong part of 1PE is its tensor force, which does not contribute in singlet states
where only the (weak) central force has a presence. The momentum-space 1PE includes also a constant term/contact term 
[see  Eq.~(\ref{eq_1pe_alt})],
which converts into a $\delta(\vec r)$-function in position space. The L0 case includes
the $\delta(\vec r)$-function contribution.

We now turn to the frame on the right of Fig.~\ref{1s0-wave}, where the 2PE exchanges of the various orders are added in.
The NL$\Omega$ 2PE (curve NL2) does not create any improvement over the L2 case. 
However 2PE at NNL$\Omega$ (curve NNL2) 
leads to an excellent reproduction of the $^1S_0$ phase shifts up to 300 MeV.
Adding more contacts beyond two in the cases of NL$\Omega$ and NNL$\Omega$ does not
improve the description, which is why we do not show these cases.
The NNNL4 case creates further subtle refinements.

We remind the reader again of the fact that
the chiral 2PE expressions at orders NL$\Omega$ and NNL$\Omega$
include polynomial terms of order $Q^0$ and $Q^2$,
and the 2PE expressions at order N$^3$L$\Omega$
include polynomial terms up to order $Q^4$, which are always
compensated by contacts of the same order.
Therefore, in the case of the $^1S_0$ wave, the minimal number of contacts to be applied at
 NL$\Omega$ and NNL$\Omega$ is two (zeroth and second order) and four (of orders zero, two, and four) at N$^3$L$\Omega$.
Thus, the contact LECs shown in Table~\ref{LEC_1s0-wave} for  NL2, NNL2, and NNNL4
are renormalized numbers whose size does not necessarily reflect the size of what is missing
beyond the genuine pion exchange contributions.

In summary, contacts alone are inadequate to describe the $^1S_0$-wave at  intermediate energies. 
The strong chiral 2PE that starts at NNL$\Omega$ is needed for a quantitative description of the $^1S_0$-wave.
There is a clear signature of chiral symmetry in $^1S_0$-wave.

\subsection{The coupled $^3S_1$-$^3D_1$-$\epsilon_1$ system}

\begin{table}
\caption{Columns two to seven show the contact LECs used in the $^3S_1-^3D_1$ waves
[cf.\ Eq.~(\ref{eq_ct})].
 The  $\widetilde{C}_\alpha$ of the zeroth order contact are given in 
 units of $10^4$ GeV$^{-2}$;
the $C_\alpha$ of second order in $10^4$ GeV$^{-4}$; 
and $\widehat{D}_\alpha$ and $D_\alpha$ of fourth order
 in $10^4$ GeV$^{-6}$.
 Column eight and nine display the triplet scattering length, $a_{t}$, and effective range, $r_{t}$, respectively, in the ${{}^{3}S_1}$ state.  }
 \begin{tabular*}{\textwidth}{@{\extracolsep{\fill}}lllllllll}
    \hline \hline
       \noalign{\smallskip}
   case & $\widetilde{C}_{{}^{3}S_1}$ &  $C_{{}^{3}S_1}$ & $\widehat{D}_{{}^{3}S_1}$ & $D_{{}^{3}S_1}$ & ${D}_{{}^{3}D_1}$ & $C_{{{}^{3}S_1}-{{}^{3}D_1}}$ & $a_t$ (fm) & $r_t$ (fm) \\  
    \noalign{\smallskip}
     \hline
       \noalign{\smallskip}
        ct1 & -0.077103 & 0 & 0 & 0 & 0 & 0 & 5.42 & 0.68 \\
    \noalign{\smallskip}
        ct5 & -0.1311 & 2.0 & -0.5 & 0 & 27.0 & -1.25 & 5.42 & 1.76 \\
     \noalign{\smallskip}
    L1 & -0.06366 & 0 & 0 & 0 & 0 & 0 & 5.42 & 1.59 \\
   \noalign{\smallskip}
   L5 & -0.13345 &0.4 & -0.7 & 0 & -2.0 & 0.41 & 5.42 & 1.73 \\
   \noalign{\smallskip}
   NL2 & -0.136835 & -0.39 & 0 & 0 & 0 &0 & 5.42 & 1.76 \\
         \noalign{\smallskip}
    NL5 & -0.1255 & -0.5 & -2.3 & 0 & -2.3 & 0.1 & 5.42 & 1.73 \\
            \noalign{\smallskip}
    NNL2 & -0.10002 & -0.335 & 0 & 0 & 0 & 0 & 5.42 & 1.75 \\
               \noalign{\smallskip}
    NNL5 & -0.14875 & 0.4 & -0.1 & 0 & -1.4 & 0.4 & 5.42 & 1.74 \\
                \noalign{\smallskip}
    NNNL8$^a$ & -0.159635 & 0.8233 & -4.319 & -19.17 & -5.59 & 0.503 & 5.42 & 1.75 \\
        \noalign{\smallskip}
       \hline\hline
       \end{tabular*}
       \footnotesize
       $^a$ In the case of NNNL8, besides the six parameters given, 
       $\widehat{D}_{^3S_1-^3D_1}=1.162$ and $D_{^3S_1-^3D_1}=1.759$. In all other cases,
       $\widehat{D}_{^3S_1-^3D_1}=D_{^3S_1-^3D_1}=0$.
  \label{LEC_SD-waves}
\end{table}

\begin{figure}[t]\centering
\vspace{-1.5cm}
\subfloat{\includegraphics[width = 2.4 in]{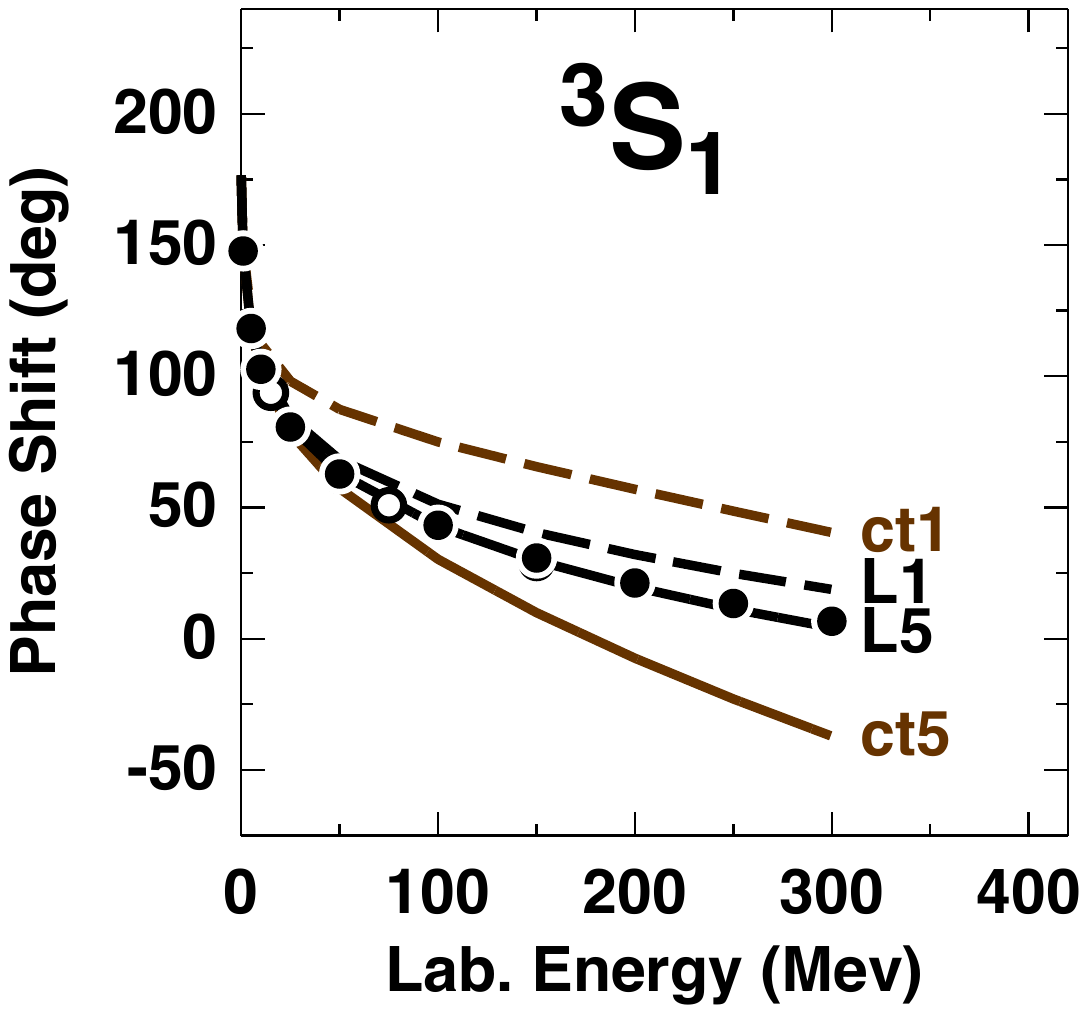}}
\subfloat{\includegraphics[width = 2.4 in]{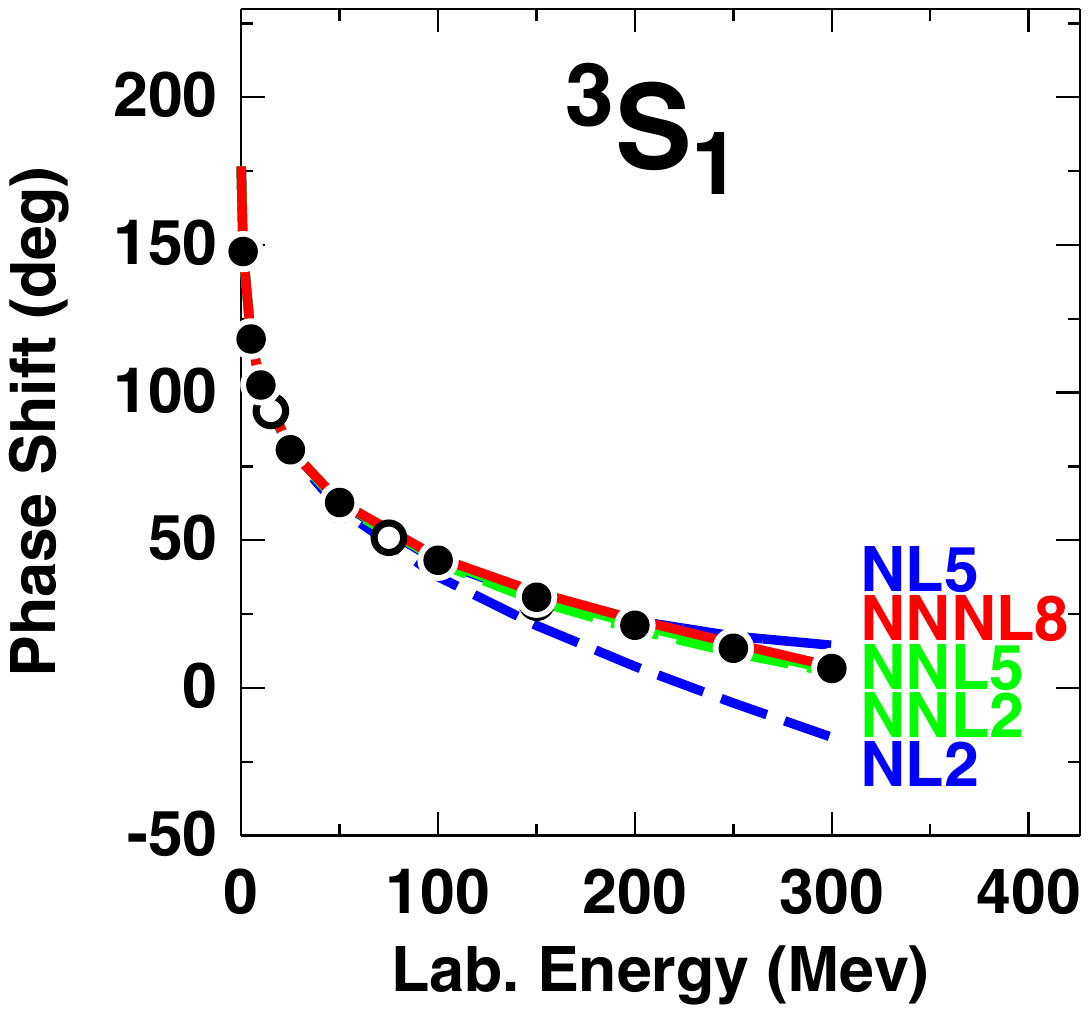}}
\\
\vspace{-2.5cm}
\subfloat{\includegraphics[width = 2.4 in]{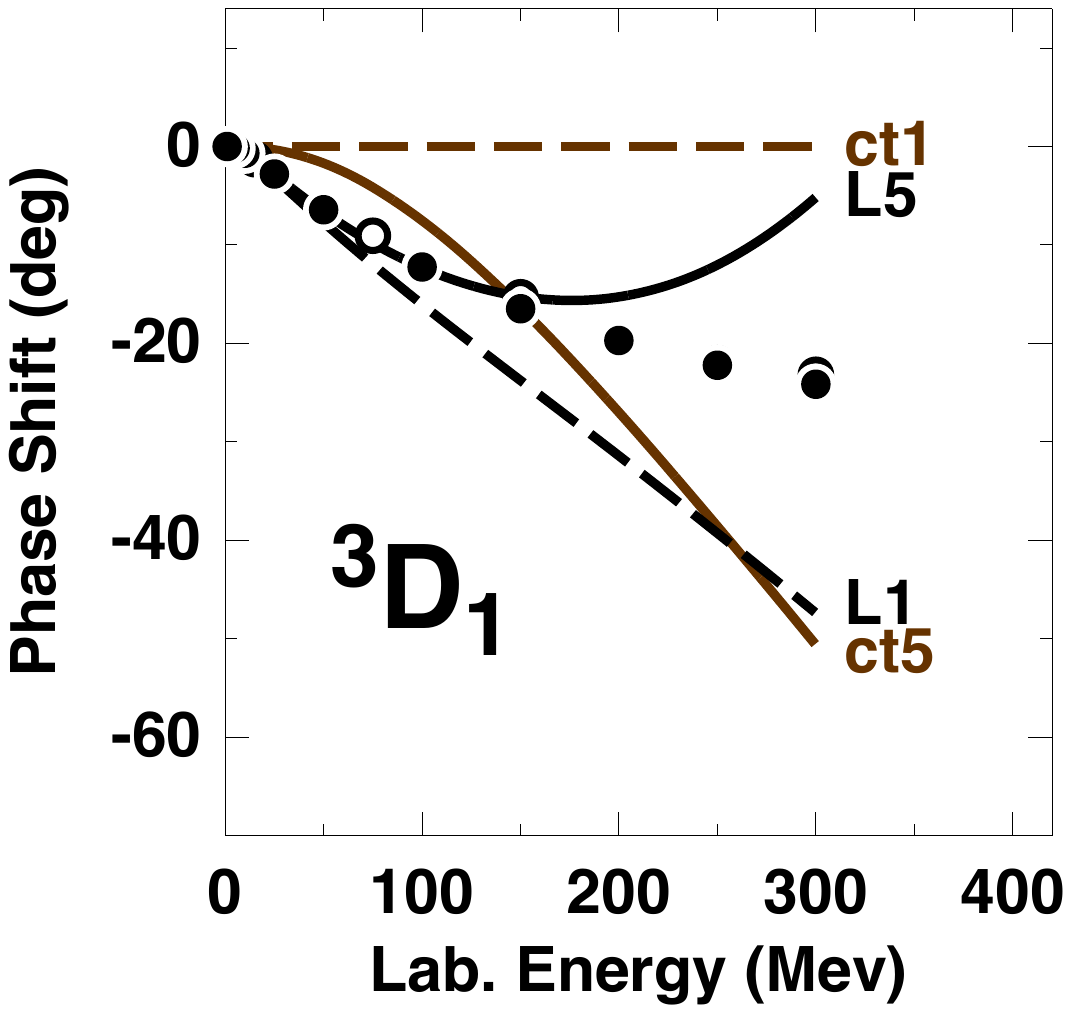}}
\subfloat{\includegraphics[width = 2.4 in]{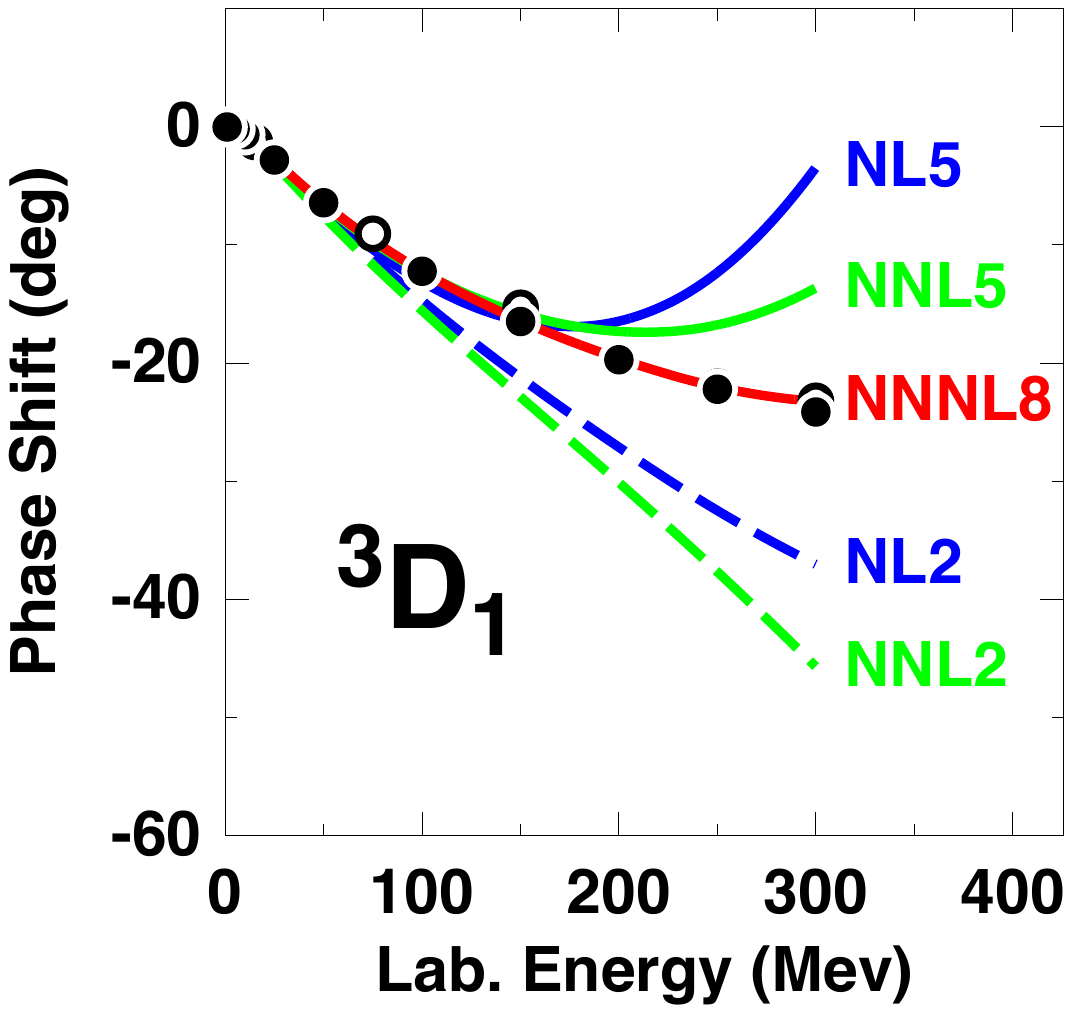}}
\\
\vspace{-2.5cm}
\subfloat{\includegraphics[width = 2.4 in]{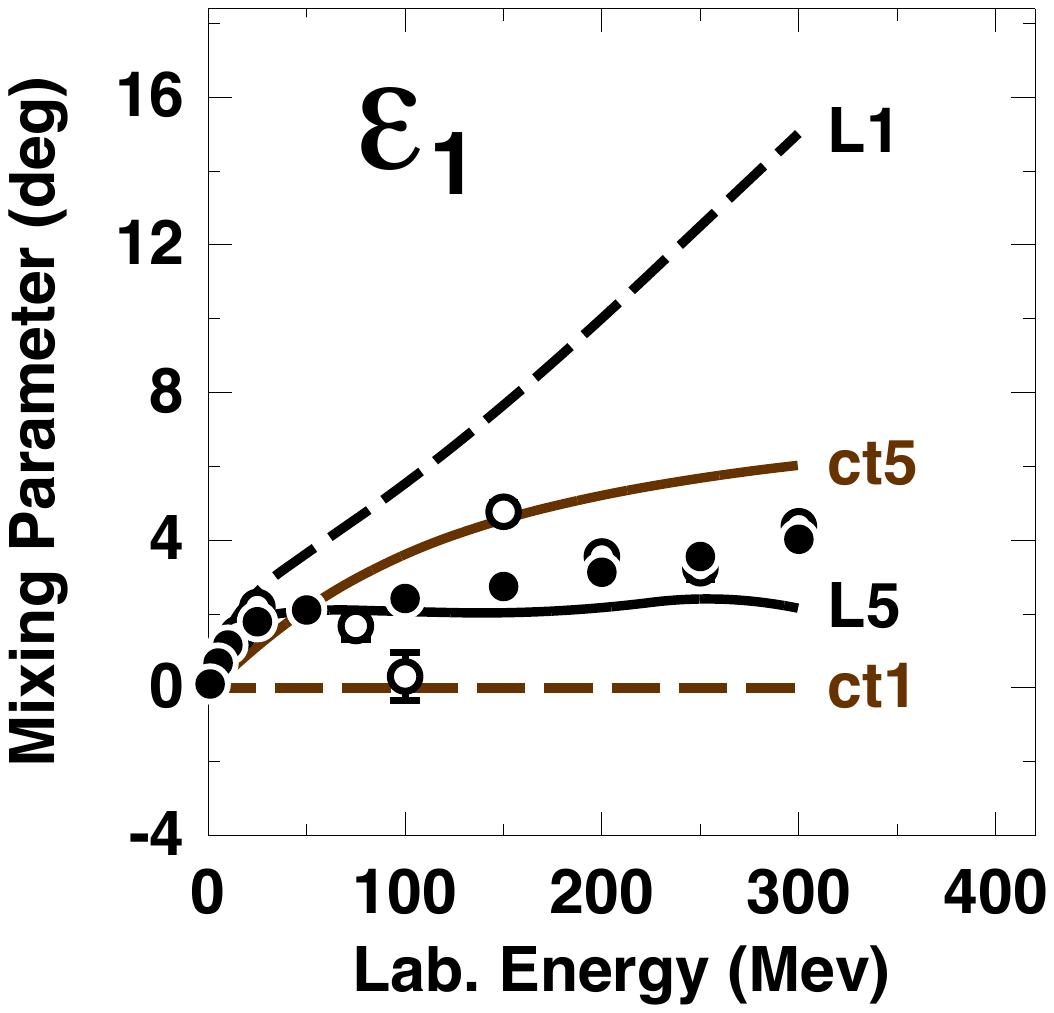}}
\subfloat{\includegraphics[width = 2.4 in]{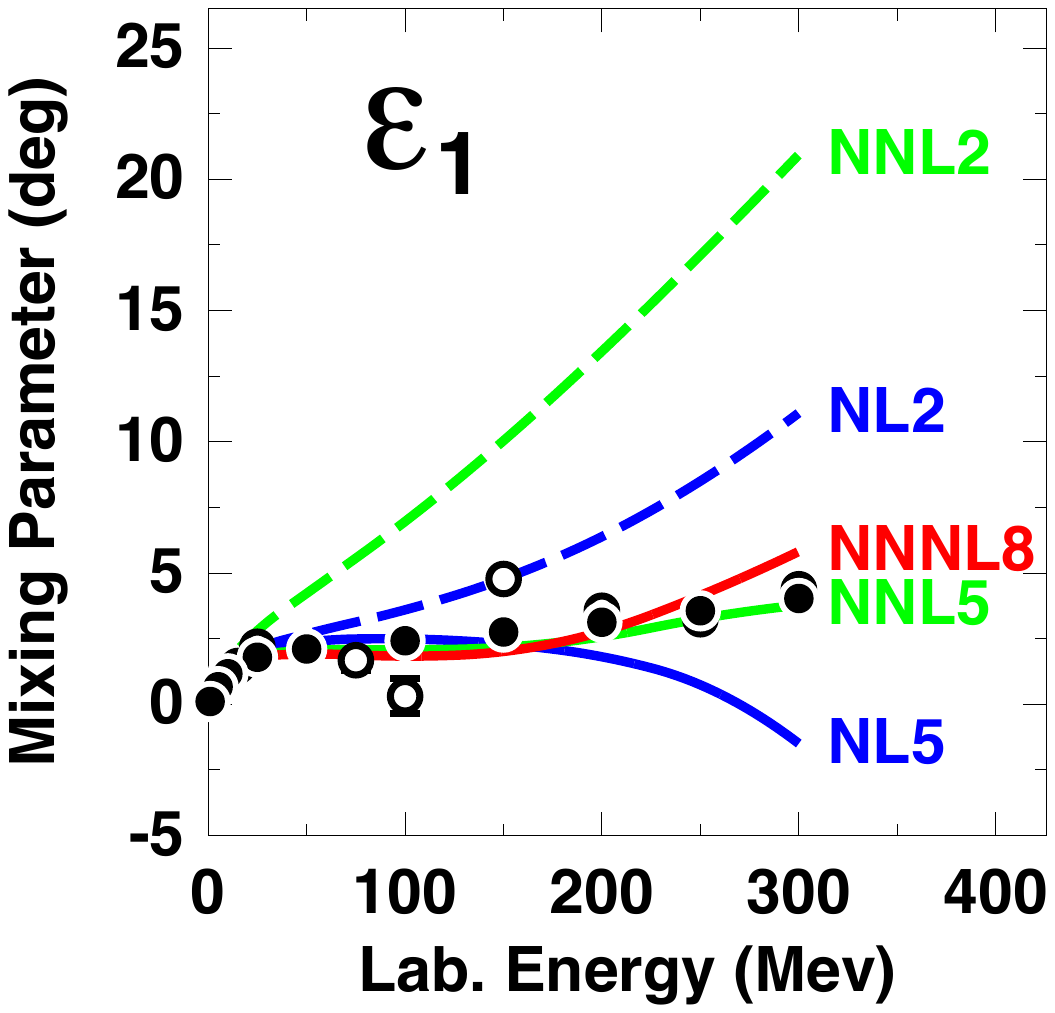}}
\vspace{-1.5cm}
\caption{$^3S_1$, $^3D_1$, and $\epsilon_1$ phase parameters of neutron-proton scattering 
for the various cases discussed in the text.
Solid dots and open circles as in Fig.~\ref{fig_ph1a}.}
\label{SD-waves}
\end{figure}

In the coupled $^3S_1$-$^3D_1$-$\epsilon_1$ system, we have available a total of eight contact terms [cf.\ Eq.~(\ref{eq_ct})]; namely, 
four for $^3S_1$ ($\widetilde{C}_{{}^{3}S_1}$, $C_{{}^{3}S_1}$, $\widehat{D}_{{}^{3}S_1}$, $D_{{}^{3}S_1}$),
one for $^3D_1$ (${D}_{{}^{3}D_1}$), and three for the ${{}^{3}S_1}$-${{}^{3}D_1}$
transition potential ($C_{{{}^{3}S_1}-{{}^{3}D_1}}$,
$\widehat{D}_{{{}^{3}S_1}-{{}^{3}D_1}}$, ${D}_{{{}^{3}S_1}-{{}^{3}D_1}}$).
When we use only one of the eight contacts, we pick the zeroth order one, $\widetilde{C}_{{}^{3}S_1}$, and fit it to the
$^3S_1$ scattering length, $a_t=5.42$ MeV. When we apply the two contacts
$\widetilde{C}_{^3S_1}$ and $C_{^3S_1}$,
we fit, besides the scattering length, the $^3S_1$ effective range parameter, 
$r_t=1.75\pm 0.02$ MeV. Using the three $^3S_1$ parameters
$\widetilde{C}_{^3S_1}$, $C_{^3S_1}$, and $\widehat{D}_{^3S_1}$,
  we try to also reproduce (if possible) the
empirical $^3S_1$ phase-shift at 50 MeV laboratory energy as determined in the Nijmegen phase-shift analysis~\cite{Sto93}.
Besides the three contact LECs mentioned, we will, in some cases, also include $D_{^3D_1}$ and
$C_{^3S_1-^3D_1}$, which affect the $^3D_1$ phase shift and the $\epsilon_1$ parameter,
respectively.
To prevent our investigation from becoming too involved, we do not vary the LECs
$D_{^3S_1}$, $\widehat{D}_{^3S_1-^3D_1}$, and $D_{^3S_1-^3D_1}$ at orders
up to NNL$\Omega$ and keep them at zero.
Thus, up to NNL$\Omega$, we will be experimenting with maximally five contacts in the  $^3S_1$-$^3D_1$-$\epsilon_1$ system.

We consider the following cases with
the short notation given in parenthesis.
\begin{itemize}
\item
One contact contribution, and nothing else (ct1). 
\item
Five contact contributions (ct5). 
\item
L$\Omega$ pion exchange (i.~e., 1PE) plus one contact term (L1).
\item
L$\Omega$ 1PE plus five contact terms (L5).
\item
NL$\Omega$ pion exchanges plus two contact terms (NL2).
\item
NL$\Omega$ pion exchanges plus five contact terms (NL5).
\item
NNL$\Omega$ pion exchanges plus two contact terms (NNL2).
\item
NNL$\Omega$ pion exchanges plus five contact terms (NNL5).
\item
NNNL$\Omega$ pion exchanges plus eight contact terms (NNNL8).
\end{itemize}

The values for the contact LECs used in the various cases are listed in 
Table~\ref{LEC_SD-waves}, and
the corresponding phase shifts up to 300 MeV are
shown in Fig.~\ref{SD-waves}.

When only one contact term is used (fit to $a_t$) and no pion contributions (case ct1), 
then the 
$^3S_1$ phase shifts at  intermediate energies are substantially above the data and $r_t$ is off by about 1 fm. 
Adding one more contact (case ct2, not shown), gets $r_t$ correct, but moves the phase shifts 
at intermediate energies far below the data, very similar to the case ct5 that is shown in
Fig.~\ref{SD-waves}. In fact, adding more contacts to the coupled system under consideration does not change the $^3S_1$ phase shifts up to the maximum of five contacts.
 Clearly, contacts alone cannot describe the $^3S_1$ wave at  intermediate energies, no matter how many contacts one is using. However, adding 1PE (case L1) makes a big difference, getting the $^3S_1$
 phase shifts almost right and finally perfect with more contacts (L5).
 This is quite in contrast to $^1S_0$, where 1PE has little influence and where 1PE plus contacts never lead to a reproduction of the phase shifts. The reason for this is that, in the coupled
 $^3S_1$-$^3D_1$ state, the 1PE tensor force contributes strongly which is crucial for
 the correct description of this coupled system.

We now turn to the $^3S_1$ frame on the right of Fig.~\ref{SD-waves}, where the 2PE exchanges of the various orders are added, and we see that 2PE does not make much difference.

Turning to the $^3D_1$ phase shifts, we see again that contacts alone cannot get this partial wave right. The contact contribution is too short-ranged for this partial wave as clearly seen by the very small contribution at low energies and too strong a contribution above 150 MeV. Adding 1PE
gets it right at low energies, but requires short-ranged corrections at higher energies. This can be done by contacts (case L5) or by 2PE contributions of higher order together with moderate
contacts (right $^3D_1$ frame).

Finally, we turn to the $\epsilon_1$ parameter, which is interesting, because it is proportional
to the $^3S_1$-$^3D_1$ transition potential created exclusively by the tensor force.
1PE generates a (too) strong tensor force (L1) which, when damped by a short ranged contact, gets it about right (L5). The 2PE of the various orders do also generate more or less tensor force contributions which require short-range contact corrections to get it right.

Thus, qualitatively, 1PE plus a short-range correction is all that is needed for the $^3S_1$-$^3D_1$ system. Interestingly, the chiral 2PE contributions are not important in this case. The deeper reason for this is that the iteration of the 1PE tensor force in this coupled system
generates a 2PE contribution that is so strong  that it makes other 2PE contributions insignificant.

Because of the polynomial terms that accompany chiral 2PE contributions,
in the case of the coupled $^3S_1$-$^3D_1$ system, the minimal number of contacts to be applied at
 NL$\Omega$ and NNL$\Omega$ is three, namely, 
 $\widetilde{C}_{^3S_1}$, $C_{^3S_1}$, and $C_{^3S_1-^3D_1}$.
  In the case of N$^3$L$\Omega$ it is eight.

In summary, contacts alone are inadequate to describe the 
$^3S_1$-$^3D_1$-$\epsilon_1$ system at  intermediate energies.
Crucial is the 1PE which, for good reasons, is called the {\it Leading Order} of the chiral expansion.


\section{Summary and conclusions}
\label{sec_sum}

The most characteristic feature
in the design of chiral $NN$ potentials is that the long-range part of the 
potential is described by
one- and multi-pion exchanges which are ruled by chiral symmetry.
In contrast, the short-range part consists
 simply of polynomial terms (``contact'' terms),
since the short-range nucleon structure cannot be resolved at low energies.

In the lower partial waves of $NN$ scattering,
which are the dominant ones for predictions of observables of nuclear structure
and reactions, 
  contacts as well as pion-exchanges contribute. 
But, since lower waves are more sensitive to the short-range, the contacts may be 
squashing the pion-exchange contributions, thus, diminishing the role of 
chiral symmetry for those predictions.

Hence, the purpose of this study was to investigate the role of the contacts, on the one hand, and the effect of the pion exchanges, on the other hand, in the lower partial waves of chiral $NN$ potentials. 
 
We have shown in detail, what contact terms alone can achieve.
 This is displayed by the brown ct curves in the
left frames of Figs.~\ref{D-waves} to \ref{SD-waves}, which all demonstrate that
contacts alone are totally inadequate and do not catch anything of the nature of the nuclear
force in those partial waves.
Adding (chiral) 1PE yields semi-realistic results in some specific partial-wave states, where
the tensor force plays an outstanding role. Such cases are the $^3D_2$ state and 
the $^3S_1$-$^3D_1$-$\epsilon_1$ system that is coupled through the tensor force.
Chiral 2PE at NL$\Omega$ is generally weak and, therefore, does not bring about much improvement.
However, the NNL$\Omega$ 2PE is strong, creating a realistic
intermediate range attraction that cannot be simulated by contacts.

This fact is also reflected in the $\chi^2$ calculations for the fit of the $NN$ data
conducted in Ref.~\cite{EMN17}.
While the $\chi^2$/datum at NL$\Omega$ comes out to be 51.5, at NNL$\Omega$ it is
6.3, even though in both cases the number of contact terms is the same. The improvement
in the $\chi^2$ is due to an improvement of the chiral 2PE at NNL$\Omega$. Obviously, the contacts cannot substitute the chiral terms.

For very low energies, the so-called {\it pionless} EFT has been developed~\cite{HKK20},
which consists only of contact terms and does not include any pion-exchange contributions.
In view of our rather poor results for most of our ``contacts only'' fits,
one may wonder, how well the pionless EFT is doing in describing $NN$ scattering.
It needs to be explained that the pionless EFT is meant to be used only for momenta
less than the pion mass, say, 100 MeV/c CMS momentum or less, which is
equivalent to a laboratory energy of about 20 MeV. Moreover, the pionless
theory is mostly used only for $S$-waves. Fitting the $S$-waves at low energies
is then understood as reproducing the two effective range parameters.
As shown in Table~\ref{LEC_1s0-wave}, cases ct2 and ct3, 
the $^1S_0$ $a_{np}$ and $r_{np}$
can be reproduced o.k.\ with two or three contacts.
Concerning $^3S_1$, the ct5 case shown in
Table~\ref{LEC_SD-waves} demonstrates a perfect description of $a_t$ and $r_t$.
Note that for just two contacts in $^3S_1$, the same result is obtained (not shown in the table). Also the $^1S_0$ and $^3S_1$ phase shifts up to 25 MeV (or even 50 MeV) are 
reasonably close to the empirical ones, 
Figs.~\ref{1s0-wave} and \ref{SD-waves}.
But, what our fits also show is that, when one moves above about 50 MeV,
contacts only are inadequate and pion-contributions are needed
in a decisive way.

In this context, it should be noted that, in the pionless theory, better fits above 50 MeV can 
probably be achieved if one allows, e.~g., for different parametrizations in different spin-isospin channels. In the present work, we have not attempted this, but see, e.~g., 
Ref.~\cite{CRS99}.

Finally, we also note that fitting phase shifts is not everything. The ultimate purpose of  nuclear forces (in the context of nuclear structure) is to bind nuclei (with the proper binding energies). Thus, to judge the failure or sucess of different contributions to nuclear forces,
it would be interesting to study what (combination of) contributions are needed to bind nuclei properly,
where the analysis should be subdivided into the consideration of light, intermediate, and heavy nuclei. First attempts to examine light nuclei in terms of pionless forces
have been started~\cite{HKK20}, but so far there have not been any comprehensive
inquiries. This subject represents a very attractive topic for future research.

In conclusion, despite the fact that contact and pion-exchange contributions are entangled in
the all important lower partial waves of an $NN$ potential,
we were able to disentangle them.
We managed to identify and pin down
many characteristic signatures of chiral symmetry that are crucial for
the quantitaive description of the nuclear force in those low angular momentum states. 
However, that does not imply that contacts are totally useless.
For the accurate fit of $NN$ quantities, like, the effective range parameters,
the phase shifts at low energies, and the deuteron binding energy, contacts are needed. They play a subtle role and
are like the ``dot over the i''.

\begin{acknowledgements}
This work was supported in part by the U.S. Department of Energy
under Grant No.~DE-FG02-03ER41270.
\end{acknowledgements}

\end{document}